\documentclass[final]{elsarticle}
\usepackage{amsmath}
 
\usepackage{lineno,hyperref}
\usepackage[dvipsnames]{xcolor}
\usepackage{subcaption}
\usepackage{xcolor}
\usepackage[utf8x]{inputenc}
\usepackage{chngcntr}
\usepackage{hyperref}
\usepackage[markup=underlined]{changes}
\counterwithout{figure}{subsection}
\usepackage{changes}
\journal{Nuclear Inst. and Methods in Physics Research, A}

\makeatletter
\def\ps@pprintTitle{
 \def\@oddfoot{}%
 \let\@evenfoot\@oddfoot}
\makeatother

\begin{document}


\begin{frontmatter}

\title{Neutron Tomography of Spent Fuel Casks}



\author[mymainaddress]{Zhihua Liu\corref{mycorrespondingauthor}}
\cortext[mycorrespondingauthor]{Corresponding author. Tel.: 2179043453}
\ead{zhihual2@illinois.edu}

\author[mymainaddress]{Ming Fang}
\author[mysecondaryaddress]{Jon George}
\author[mymainaddress,mythirdaddress,myforthaddress]{Ling-Jian Meng}
\author[mymainaddress]{Angela Di Fulvio}

\address[mymainaddress]{Department of Nuclear, Plasma, and Radiological
                        Engineering, \\University of Illinois, Urbana-Champaign,
                        \\ 104 South Wright Street, Urbana, IL 61801, United States}
\address[mysecondaryaddress]{Department of Radiation and Cellular Oncology, \\University of Chicago Medicine,
                        \\ 5758 S Maryland Ave, Chicago, IL 60637, United States}
\address[mythirdaddress]{Beckman Institute for Advance Science and 
                      Technology, \\University of Illinois, Urbana-Champaign,
                        \\ 405 N Mathews Ave, Urbana, IL 61801, United States} 
\address[myforthaddress]{Department of Bioengineering, \\University of Illinois,
                        Urbana-Champaign, \\ 1406 W Green St, Urbana, IL 61801, 
                        United States}

\begin{abstract}
    Dry casks for spent nuclear fuel (SNF) ensure the safe storage of SNF and provide radiation shielding. However, the presence of the thick casks encompassing several layers of steel and concrete makes inspection of the SNF a challenging task. 
    Fast neutron interrogation is a viable method for the nondestructive assay of dry storage casks. In this study, we performed a Monte Carlo \replaced{simulation-based}{simulation based} study associated with \replaced{a machine-learning-based}{an machine-learning based} image reconstruction method to verify the content of SNF dry storage casks. We studied the use of neutron transmission and back-scattered measurements to assess the potential damage to fuel assemblies or fuel pin diversion during transportation of dry casks. We used Geant4 to model a realistic HI-STAR 100 cask, MPC-68 canister and basket, and GE-14 fuel assembly irradiated by a D-T neutron generator. Several bundle diversion scenarios were simulated. The angular distribution of the neutrons scattered by the \replaced{}{the}cask was used to identify the diversions inside the fuel cask. \added{A fuel bundle with at least 75\% of its pins removed can be identified with a drop in the back-scattered signature larger than $2\sigma$ compared with a fully loaded scenario.} \replaced{}{We applied a\added{n} iterative linear inverse approach to reconstruct the cask loading and obtained well reconstructed images for the peripheral assemblies. In this configuration  }
    \replaced{}{a fuel bundle with at least 75\% of the pins removed can be identified from a drop in the neutron counts detected in the 20$^o$-30$^o$ interval.} \added{We combined an iterative reconstruction algorithm with a convolutional neural network (CNN) to obtain a cross-sectional image of the fuel inside the cask.} \replaced{}{and improve the quality of the reconstructed images. The signal-to-noise ratio (SNR) can be improved by a factor of at least 1.5.} The proposed imaging approach allows \replaced{locating}{to locate} the position of a missing fuel bundle \added{with at least 75\% of the pins removed} when performing \replaced{}{a} \replaced{tomographic imaging of a canister with an overall scan time of less than two hours}{1-hour long tomographic image with an overall scan time}, when using a commercial neutron generator with a source strength of \replaced{$10^{10}$}{$10^9$} n/s in the 4$\pi$ solid angle.

\end{abstract}


\end{frontmatter}


\section{Introduction}

Spent nuclear fuel (SNF) bundles can be stored in water pools, dry storage and transport casks, and vault-type storage facilities \cite{international2012iaea}. \replaced{By the end of 2020, there are 94 operating commercial nuclear reactors in the United States.}{Among the currently operating nuclear power plants in the United States,} Most of water pools at reactor sites have almost reached their full capacity \cite{greulich2017high}. \replaced{When the capacity of the water pool is reached or there is a need to transfer SNF to an away-from-reactor (AFR) reprocessing plant or a long-term storage site, the SNF is removed from the water pool and loaded into the interim dry storage cask or transfer cask before transportation.
Therefore, the application of the dry cask storage and transport service is becoming more and more important with the increasing need to transfer SNF bundles from an at-reactor (AR) pool to an AFR storage \cite{NRCwebsite1}.}{When the capacity of the water pool is reached or there's a need to transfer SNF to an AFR reprocessing plant or long-term storage site, the SNF are removed from the water pool and loaded 
into the interim dry storage cask or transfer cask before transportation.
Therefore, the need to transfer SNF bundles from an AR pool to an AFR storage is continuously increasing \cite{NRCwebsite1}, which makes the application of the dry cask in the dry storage and transport service of SNF bundles very important.} Verifying the integrity of the SNF without reopening the sealed containers or dry casks is an open technological challenge. 

Continuity of knowledge (CoK) to ensure that the information on spent fuel is uninterrupted and authentic \added{\cite{blair2014global}} 
is typically ensured using seals and visual observation. However, some cases may require reestablishment of inventory after the loss of CoK. Currently, there are no generally accepted or fully verified methods to reestablish the CoK without reopening the dry casks. 
Therefore, a reliable method to non-destructively assay SNF casks and verify their content and integrity is needed.

There have been several approaches proposed for the verification and detection of SNF dry casks. These approaches are mostly simulation-based studies and some have been extended to experimental verification. Scientists at Idaho National Laboratory (INL) developed the Compton Dry-Cask Imaging Scanner (CDCIS) which \replaced{identified}{identifies} the presence of SNF based on passive gamma rays emitted by SNF \cite{wharton2013summary}. The CDCIS \replaced{}{worked in a test measurement with a MC-10 cask but} failed when tested at Doel, Belgium nuclear power plant due to the thick ballistic shield on top of the Belgian cask \cite{wharton2013summary}. Cosmic-ray muon computed tomography is another technique that has been both theoretically and experimentally evaluated. Researchers from Los Alamos National Laboratory (LANL) and the University of New Mexico applied filtered back projection to reconstruct images using muon scattering and attenuation signatures, respectively \cite{poulson2017cosmic}. Researchers from LANL, the University of New Mexico, and INL measured MC-10 cask using cosmic-ray muon computed tomography \cite{durham2018verification}. Researchers from Oak Ridge National Laboratory conducted a study to image the dry storage casks using a simple Point of Closest Approach (PoCA) algorithm with muons \cite{chatzidakis2016investigation}\cite{chatzidakis2017classification}. Other approaches include using Thomson-scattering quasi-monoenergetic photon sources on a MC-10 cask that can be applied to MPC-24, MPC-32 and MPC-68 casks \cite{miller2020verification}, using high-energy X-ray computed tomography \cite{liu2016simulation}as well as high-energy neutron transmission analysis \cite{greulich2017high}.
Among these proposed verification methods, the most promising one seems to be the cosmic-ray muon tomography method. However, it requires \replaced{a long inspection time}{long inspection times} of more than one day and thus it cannot be used to verify the cask's integrity in case of an accident occurring during transportation.


In this paper, we present a method based on 14.1-MeV neutron tomography. This method aims at identifying potential fuel diversion scenarios. We exploit the penetrating power of 14.1 MeV DT neutrons together with reconstruction algorithms to improve the method's sensitivity.
First, we studied the sensitivity of the inspection method to missing assemblies. We studied the signatures that showed the highest correlations with the missing SNF assembly. Then, we used the identified signatures to carry out the neutron tomography technique and obtain the reconstructed cross-sectional images of the canister. The results showed that for an inspection consisting of 12 scans, the total measurement time is \replaced{less than 2 hours considering an intrinsic detector efficiency of 60\%}{about one hour}.


\section{Methods}


We simulated a realistic dry cask system, which consists of a HI-STAR-100 cask, MPC-68 canister and basket, and GE-14 fuel assembly using Geant4 toolkit \cite{agostinelli2003geant4}. In this section, we introduce \replaced{the simulated geometry of the cask and the computational methods used to locate missing fuel bundles.}{the detailed spent fuel dry cask {model} and the related important parameters important parameters of the model. Then we established the Geant4 model and simplified the model to improve the computational speed.} We simulated various fuel assembly diversion scenarios and obtained the neutron back-scattered signature that is related with the missing bundle locations. We used the neutron back-scattered signature and a bi-dimensional image reconstruction algorithm \added{and a CNN model }to obtain reconstructed images of the spent fuel assemblies.

\subsection{Spent fuel dry cask model}

The cask overpack is a thick  layer of stainless steel and has an additional \replaced{neutron-absorbing layer made of Holtite-A}{aluminum and boron neutron-absorbing layer} alongside the sidewall. The overall height of the cask overpack is over 5 m and its maximum diameter is approximately 2.5 m. 

The canister and basket can hold 68 boiling water reactor (BWR) fuel assemblies. Each assembly encompasses \replaced{92 SNF rods arranged in a 10$\times$10 square matrix}{10 pins by 10 pins for the SNF rods}. The assembly contains two water rods at the center, each occupying four-pin locations, and seventy full-length fuel rods, fourteen part-length fuel rods as well as eight tie rods. The full-length fuel rods are approx. 400 cm long, with approx. 380 cm of fuel pellets (the remaining length includes a plenum region and the connecting shanks), whereas the part-length fuel rods are approx. 250-cm long, with approx. 213 cm of fuel pellets \cite{proj_doc}.  The tie rods are screwed into the lower tie plate and are used to support the assembly during the operation of the fuel handling process \cite{proj_doc}.

\subsection{Geant4 model and simulation setup}

We used the Geant4 toolkit to simulate the dry cask system \cite{zhihua2018} and the interrogating source. \replaced{}{The Geant4 model was based on an AutoCAD model of the dry cask system established by collaborators in this project.} The physics list used in the Geant4 model was the QGSP\_BERT\_HP\replaced{}{ physics list}, which provides high-precision neutron transport simulation at energies below 20 MeV. 
The source term was simulated as a 14.1-MeV mono-energetic neutron beam. It was placed at the mid-plane of the cask targeting at the side surface. This energy corresponds to the energy of the neutrons produced by a Deuterium-Tritium (D-T) generator. The D-T generator produces 14.1 MeV neutrons through the following fusion reaction between deuteron ions (D) and tritium ions (T): $D + T \longrightarrow n + {^4He} + 17.6~MeV$ \cite{wiki:fusion}. The neutron carries 80\% (i.e., $14.1~MeV$) of the released energy.

\begin{figure}[!h]
\centering
\begin{tabular}[c]{ccc}
    \begin{subfigure}[c]{0.26\textwidth}
    \includegraphics[scale=0.26]{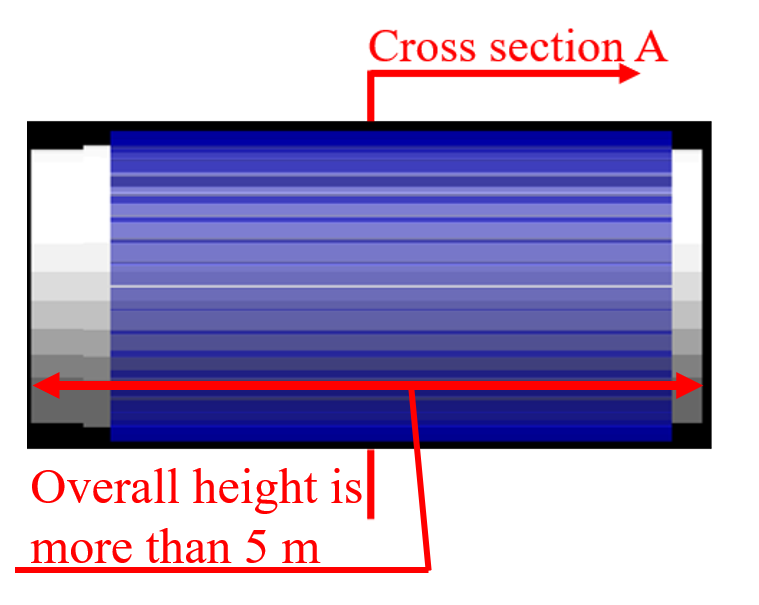}
    \caption{Side view of the cask model.}
    \label{fG4_model_view:a}
    \end{subfigure}
    
    \begin{subfigure}[c]{0.33\textwidth}
    \centering
    \includegraphics[scale=0.18]{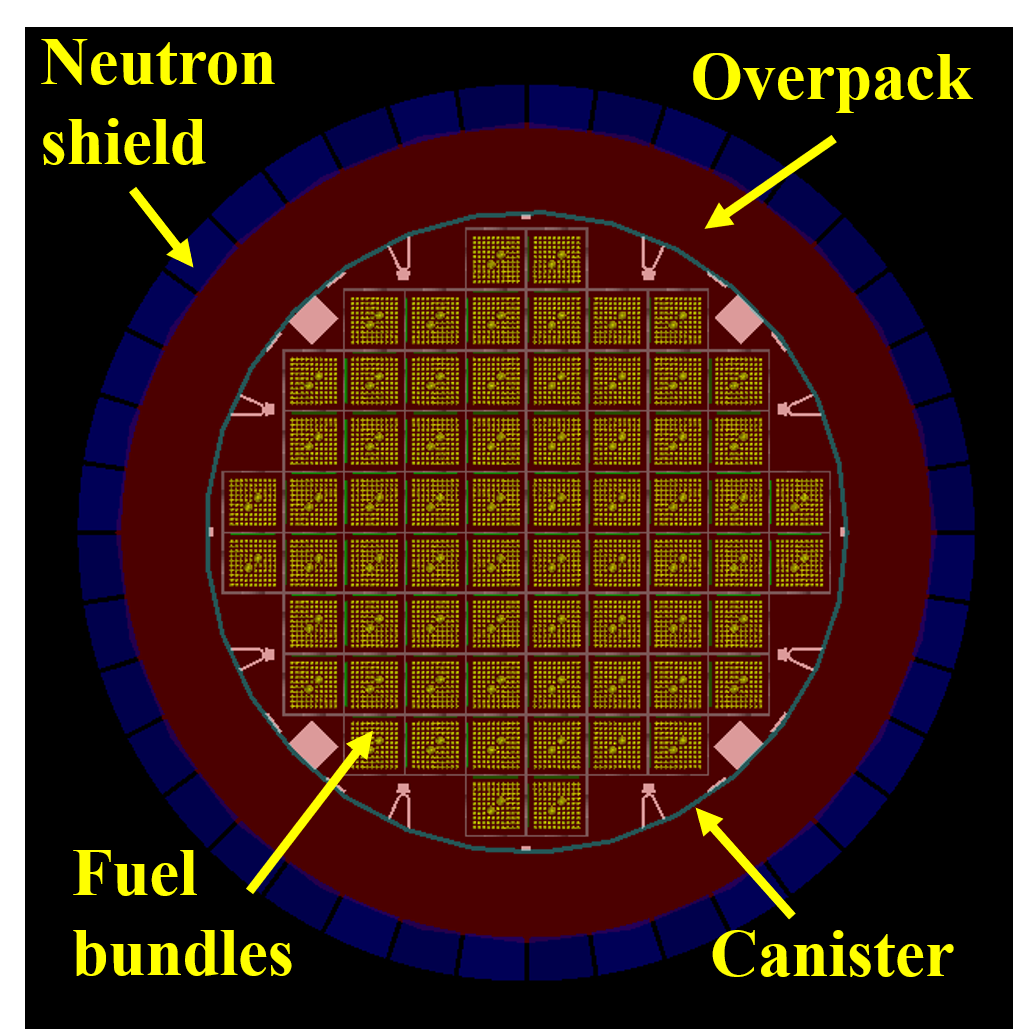}
    \caption{Cross-section A.}
    \label{fG4_model_view:b}
    \end{subfigure}
    
    \begin{subfigure}[c]{0.33\textwidth}
    \centering
    \includegraphics[scale=0.20]{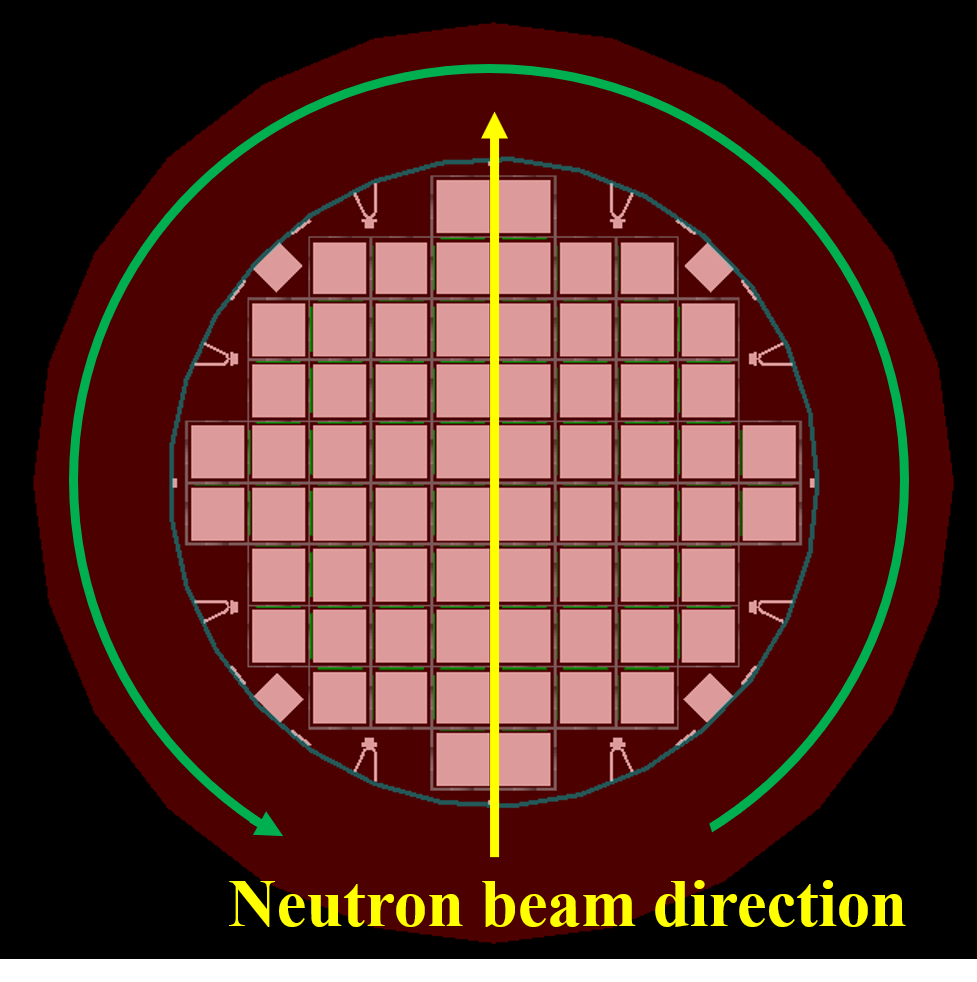}
    \caption{Homogenized Geant4 model.}
    \label{fG4_model_view:c}
    \end{subfigure}
\end{tabular}
\caption{The detailed and simplified Geant4 models.}
\label{fG4_model_view}
\end{figure}

The visualization of the Geant4 model is shown in Fig. \ref{fG4_model_view}. \added{Fig. \ref{fG4_model_view:b} shows the realistic layout of the dry cask system, including the neutron absorbing shielding layer, cask overpack, canister, fuel basket structures (honeycomb structure and neutron absorbing panel) and supports, as well as fuel bundles (fuel rods, water rods and tie rods). The orientation of the inner fuel assembly deployment is known on the outside of the dry cask sidewall.} The outermost layer shown in blue is the neutron shield layer of the cask. 
In the image reconstruction analysis, we removed both the neutron shield and the cask and kept only the canister.
\replaced{}{The reason why the neutron shield and the cask can be removed is that} During specific stages of the SNF bundles' transfer, such as moving the SNF from the water pool into the transfer cask or from the transfer cask to the final storage place, the SNF could be contained only in the canister and not surrounded by the thick metal cask and the neutron shield. Hence, the nondestructive assay (NDA) of the assemblies without neutron shield can be performed during these transfer procedures.

Fig. \ref{fG4_model_view:a} and Fig. \ref{fG4_model_view:b} show a detailed schematic of the SNF within the cask. 
We developed a simplified model with homogenized fuel assembly as shown in Fig. \ref{fG4_model_view:c} to reduce simulation run-time while keeping a realistic fuel composition. \added{The structure, compositions and layout of the simplified model are analogous to the original realistic model. The only difference between the two models is that in the homogenized composition of the fuel assembly, each assembly is modeled as a bulk volume without the geometric detail of single fuel pins.} The fuel pellets are composed of ceramic uranium dioxide and gadolinium oxide. In the simplified model, the fuel assembly contains a mixture of the same material compositions distributed evenly throughout the whole assembly volume and therefore with an overall density of 3.98 $g/cm^3$. 
While yielding similar results in terms of neutron \replaced{scattered}{scattering} and transmitted distribution, the run-time of the simplified model is approximately eight times faster than the detailed model. 


\subsection{Neutron back-scattered and transmitted signatures}
\label{neutron_signature}
We identified signatures that were highly correlated to the presence of SNF assemblies, while being minimally affected by the neutron attenuation. Inspection of spent-fuel cask is challenging because of the limited penetrability, even of high energy 14.1~MeV neutrons. The half-value layer (HVL) exerted by spent fuel to $14.1~MeV$ neutrons is 3.272 cm. With each fuel rod's diameter equal to approx. 1 cm \cite{proj_doc}, we can expect the intensity of the beam to be reduced to less than 50\% by only four fuel rods. The strong attenuation caused by the fuel assemblies will dramatically reduce the probability of detection neutrons transmitted through the canister.
We identified high-efficiency, information-rich features to address this challenge. The combination of back-scattered and transmitted neutrons was used to identify the defects in the \replaced{}{peripheral} fuel assemblies. \replaced{}{These signatures were used in a Bayesian tomographic reconstruction procedure. }

\begin{figure}[h!]
    \centering
    \captionsetup{justification=centering}
    \includegraphics[scale=0.65]{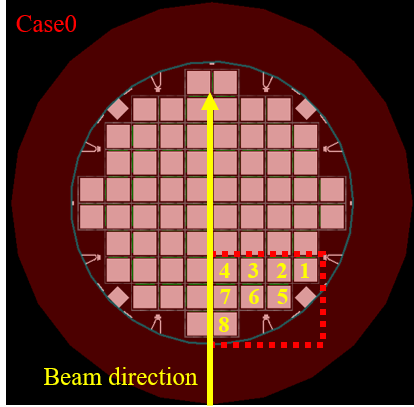}
    \caption{Several SNF assembly diversion scenarios.}
    \label{fSimulation_design}
\end{figure}

\begin{figure}[h!]
    \centering
    \captionsetup{justification=centering}
    \includegraphics[scale=0.5]{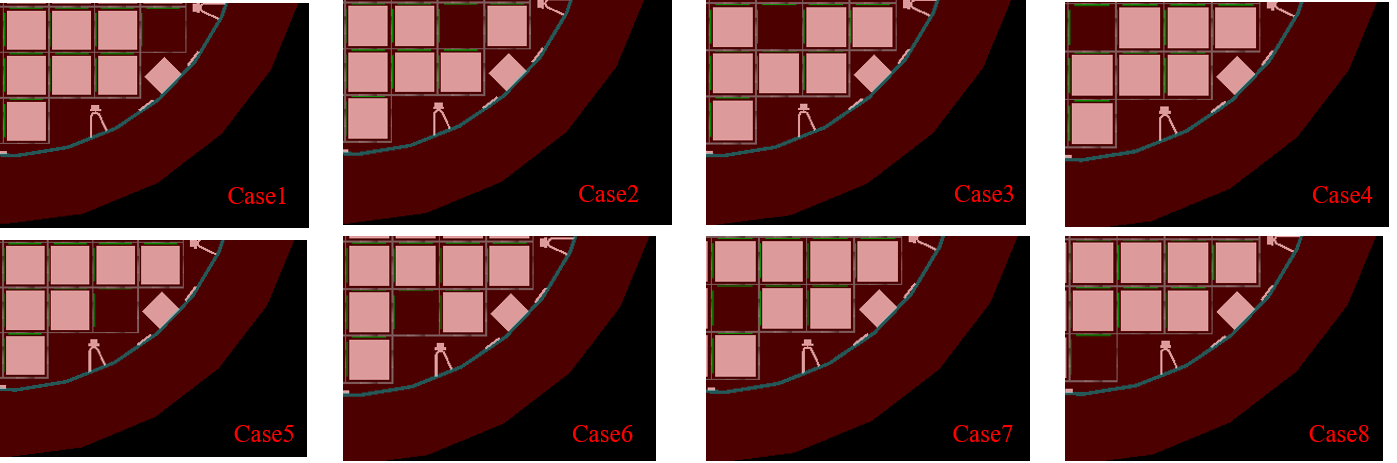}
    \caption{Eight anomalous cases for one SNF assembly missing scenario.}
    \label{fmisloading_detail}
\end{figure}

We detected the neutrons emerging from the surface of the dry cask under several fuel assembly mis-loading scenarios. \added{All neutrons that incident into the detectors are contributing to the gross neutron counts. We consider using organic scintillation detectors when estimating the actual measurement time and assume the intrinsic detection efficiency to be 60\%.} Since the fuel assembly layout is symmetric, we simulated one-quarter of the cask highlighted in Fig. \ref{fSimulation_design} within the dashed red rectangle box. The neutron beam was simulated as conical for variance reduction, and the yellow arrow in Fig. \ref{fSimulation_design} represents the neutron beam direction. We simulated eight fuel assembly anomalous scenarios and compared the corresponding results with that of a fully-loaded cask (referred to as Case0). In each anomalous scenario, we removed one fuel assembly. These scenarios are illustrated in Fig. \ref{fmisloading_detail}.

\subsection{Bi-dimensional image reconstruction through iterative algorithm}
 Spent fuel verification procedures are generally verification-based, where a previous measurement of the item is compared to the current one to ensure CoK. This procedure would greatly benefit from an imaging approach, which is more informative compared to a gross counting measurement performed outside the cask. After simulating the correlation of the back-scattered signature, as reported in Section \ref{neutron_signature}, we developed an imaging method to locate the missing bundles in a 2D cross-sectional view of the fuel \replaced{region}{bundles}. We assumed the model of the fuel bundles as a linear forward model. The linearity assumption is based on the fact that neutrons resulting from fission reactions can be neglected and therefore the problem is similar to \replaced{an}{a} X-ray tomography where the attenuation coefficient is replaced by the neutron removal macroscopic cross section.
\replaced{Traditional image reconstruction methods such as filtered back projection (FBP) derive the attenuation of radiation through an object from the length of the radiation path. FBP uses the estimated linear attenuation coefficient to form the attenuation profiles, then reconstructs images of the object through inverse Radon transform with mathematical filters applied to remove the blur from back-projection. However, due to the shielding effect of the fuel assemblies, this method needs a long measurement time to obtain transmitted signals that enable satisfactory imaging.
We used instead a linear-inverse approach solved through a sparsity-promoting algorithm, proven to yield improved imaging performance when compared to FBP \cite{fang2021quantitative}.
The distribution of fuel assemblies inside the canister can be considered as a sparse matrix if we treat the location of the missing fuel rods as 1 and other locations 0. The image reconstruction with sparse data can be achieved by solving a regularized linear inverse problem using the fast-iterative shrinkage-thresholding algorithm (FISTA) \cite{beck2009fast}.}{The image reconstruction is performed by solving a regularized linear inverse problem using the fast-iterative shrinkage-thresholding algorithm (FISTA) \cite{beck2009fast}.}

\begin{figure}[h!]
    \centering
    \captionsetup{justification=centering}
    \includegraphics[scale=0.3]{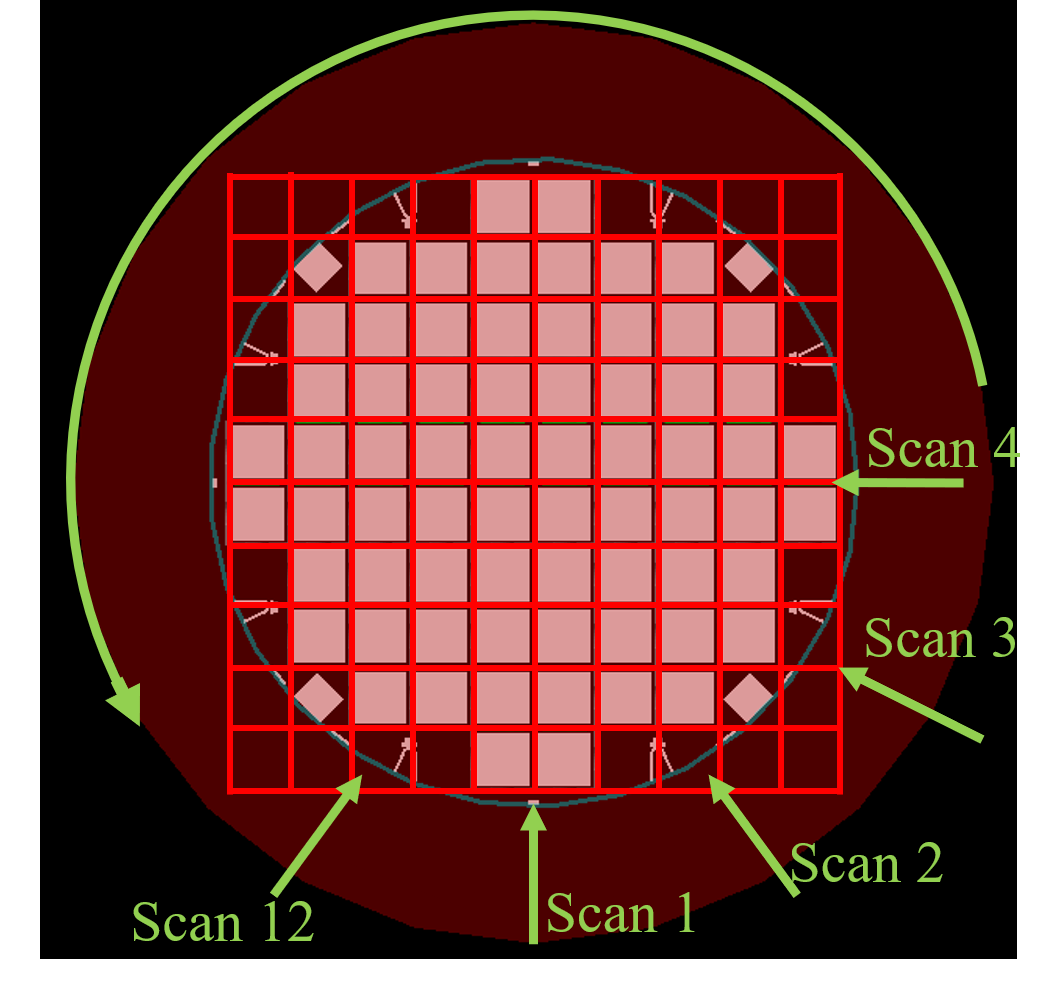}
    \caption{An example of system discretization and neutron beam direction in each scan is indicated by the arrows. }
    \label{fDiscretize}
\end{figure}  

In the inspection of the dry cask, we discretized the region-of-interest (ROI) of the fuel assembly into $N\times N$ pixels as shown in Fig. \ref{fDiscretize}, where $N=10$ in this example. Then, we performed the NDA at 12 angles around the cask, 30$^o$ apart from each other (Fig. \ref{fDiscretize}). \added{The selected number of scans is a trade-off between minimizing the relative uncertainty associated with the measurement at each angle while keeping the measurement within a reasonable time frame.} In each configuration, we removed one assembly from the fully-loaded canister and measured the back-scattered neutrons within \replaced{the selected 36 detection regions (starting from the neutron beam travelling direction, each detection region occupies $10^o$ detection angles and thus the surface of $360^o$ detection angles is divided into 36 detection regions)}{the selected two detection regions (i.e. between angle $11^o-20^o$ and $21^o-30^o$ as stated in previous section)}.  
We obtained a readout $\mathbf{y}$ with $M=12 \times 36$ entries for each missing fuel assembly scenario\added{, where 36 is the number of detection regions}. 
Therefore, the response matrix $\mathbf{A}$ of the system is a $M \times N^2$ matrix, with the column $i$ ($i=1~to~100$) being the readout of the system for the canister with the \replaced{fuel portion corresponding to the $i^{th}$ pixel removed}{$i^{th}$ fuel assembly removed}. We assume the system to be linear, i.e., the readout $\mathbf{y}$ of the canister with unknown fuel assembly removed is a linear function of the \replaced{true fuel assembly distribution $\mathbf{x}$}{intensity $\mathbf{x}$ of the true fuel assembly distribution}:
\begin{equation}
    \mathbf{y} = \mathbf{A} \mathbf{x} + \mathbf{n}
\end{equation}
where $\mathbf{n}$ stands for the random observation noise. \replaced{}{$\mathbf{y}$ and $\mathbf{A}$ can be calculated via simulation.}

Here the noise term $\mathbf{n}$ is modeled by Gaussian random noise and \replaced{ the unknown fuel assembly distribution $\mathbf{x}$} {the intensity $\mathbf{x}$ of the unknown fuel assembly distribution} is estimated by solving the following penalized least-square problem, subject to the constraint that $\mathbf{x}$ \replaced{cannot}{can not} be negative:
\begin{equation}
        \hat{\mathbf{x}} =  \underset{\mathbf{x}\geq0}{\text{argmin}} ~~ \left[\frac{1}{2}\|\mathbf{y} -\mathbf{A}\mathbf{x}\|_2^2  + \lambda \|\mathbf{x}\|_1 \right]
        \label{Eq2}
\end{equation}
In Eq. \ref{Eq2}, $\|\cdot\|_p$ denotes the $\ell_p$ norm. The first term in the function is a quadratic data fidelity term and the second term is a regularization parameter. $\lambda>0$ provides a trade-off between fidelity to the measurements and noise sensitivity. This regularization parameter promotes solutions where $\mathbf{x}$ is sparse. 
This equation can be solved using the FISTA algorithm, as detailed the following steps.

FISTA with constant step size \cite{beck2009fast}:

\textbf{Input}: $\mathbf{A},\mathbf{y}, L$: The Lipschitz constant of $\mathbf{A}, \lambda$ (depending on noise level).

\textbf{Procedure}:
\begin{enumerate}
\item  Set the initial values:

$\mathbf{z}^{(1)}= 0$, $\mathbf{x}^{(0)}= 0$, $t_1 = 1$ and $k_{\max}: total~number~of~iterations$
\item  When $1\leq k \leq k_{\max}$ is satisfied, the algorithm will do an iteration until $k_{\max}$ is reached with the following equations:
\begin{equation}
    x^{(k)}=z^{(k)}-\frac{1}{L}A^T(Az^{(k)}-\mathbf{y})
\end{equation}
\begin{equation}
    x^{(k)} =\max(x^{(k)}-\frac{\lambda}{L},0)
\end{equation}
\begin{equation}
    t_{k+1} = \frac{1+\sqrt{1+4t_k^2}}{2}
\end{equation}
\begin{equation}
    z^{(k+1)}=x^{(k)}+\frac{t_k-1}{t_{k+1}}(x^{(k)}-x^{(k-1)})
\end{equation}
\end{enumerate}
\textbf{Output}: $\hat{\mathbf{x}}$.

It will finally return the value $\hat{x} = x^{(k)}$, i.e., the reconstructed image, when the set total number of iterations is reached. 

The number of maximum iterations $k_{\max}$ is determined based on the \replaced{SNR}{signal-to-noise ratio (SNR)} of the reconstructed images and also the norm of $\|\mathbf{y} -\mathbf{A}\mathbf{x}\|$.

\subsection{Bi-dimensional Image Reconstruction through Machine Learning}

\begin{figure}[h!]
    \centering
    \captionsetup{justification=centering}
    \includegraphics[scale=0.45]{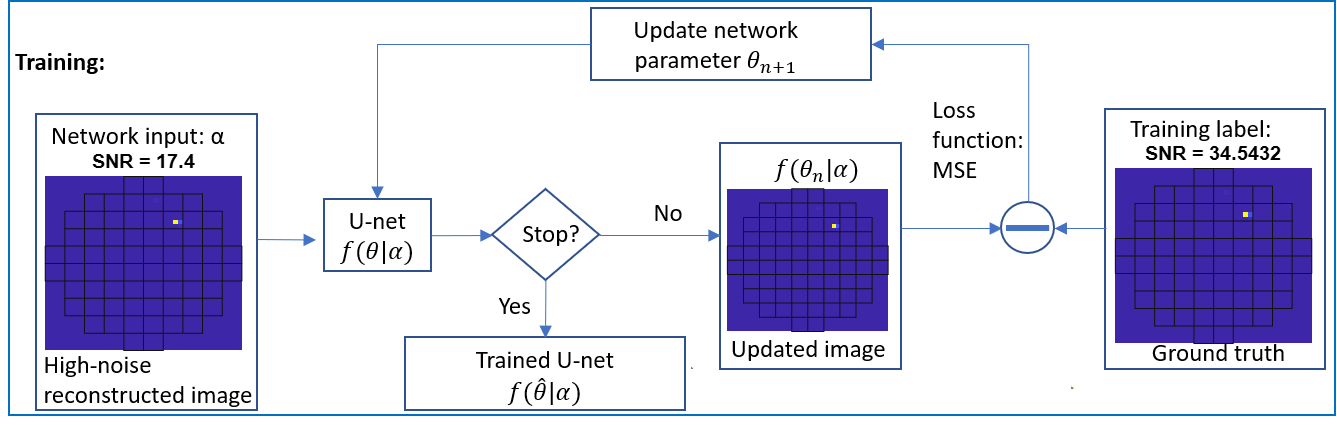}
    \caption{Block diagram of the training process of the CNN architecture.}
    \label{CNN}
\end{figure}

\added{Neural networks have been used in nuclear engineering for multiple applications, ranging from data mining \cite{korovin2015use} to radiation detection \cite{fu2018artificial}. 

We applied a CNN model to de-noise the reconstructed images obtained from FISTA to improve the image quality. Fig. \ref{CNN} shows the training process of the proposed CNN model. The CNN model is based on the 2D-Unet neural network and is pre-trained using the reconstructed images from FISTA. We use high-quality images as the training label and low-quality images as the network input. The high-quality images were generated using simulated data with low relative uncertainty of counts, and the low-quality images were generated using simulated data with few source neutrons, resulting in a high relative uncertainty of counts. The mean squared error between the network output and the training label was used a loss function of the CNN model and minimized during the training process. Based on the FISTA algorithm, the image size of the CNN input is determined as $40\times40$. The CNN featured four convolutional layers, a $3\times3$ convolutional filter and a unit batch size. The number of training epochs that minimized the loss function was selected.
}

\section{Results}

\subsection{Analysis of neutron back-scattered signature}

We detected neutrons in the $0.1 - 14.1 MeV$ range emerging from the surface of the \replaced{canister}{cask} and analyzed the sensitivity of the gross neutron count to missing fuel rods and whole assemblies as a function of the angle with respect to the interrogating beam axis. 
\replaced{}{We considered angular intervals from $0^o$ to $360^o$ with respect to the interrogating beam axis, each spanning over $10^o$ to $30^o$ (as indicated by the curved arrow shown in Fig. \ref{fG4_model_view:c}). }



\begin{figure}[h!]
    \centering
    \captionsetup{justification=centering}
    \includegraphics[scale=0.38]{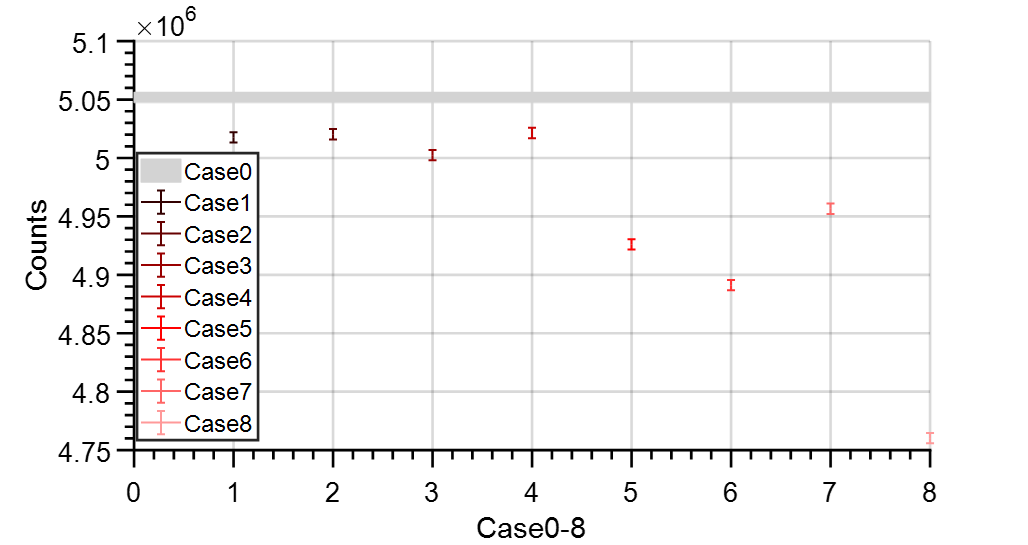}
    \caption{Back-scattered neutrons emerged from the surface of the \replaced{canister}{cask} in $1^{o}$-$10^{o}$ for different cases ($nps=10^8$).}
    \label{fnCounts_1-10}
\end{figure}

Fig. \ref{fnCounts_1-10} shows the neutron counts in the $1^o$-$10^{o}$ interval. The grey bar represents the counts detected from the fully loaded case (Case0) with an uncertainty of $\pm2\sigma$. The other eight data points indicate the counts of neutrons for the above-mentioned assembly missing cases (Case1-Case8) with an uncertainty of $\pm2\sigma$. Fig. \ref{fnCounts_1-10} shows that the \replaced{neutron}{neutrons} counts reaching the surface of the \replaced{canister}{cask} may change significantly with the number and position of the missing assemblies when compared to the fully loaded case. This signature can be used in the identification of the mis-loaded SNF assemblies.

\begin{figure}[h!]
    \centering
    \captionsetup{justification=centering}
    \includegraphics[scale=0.3]{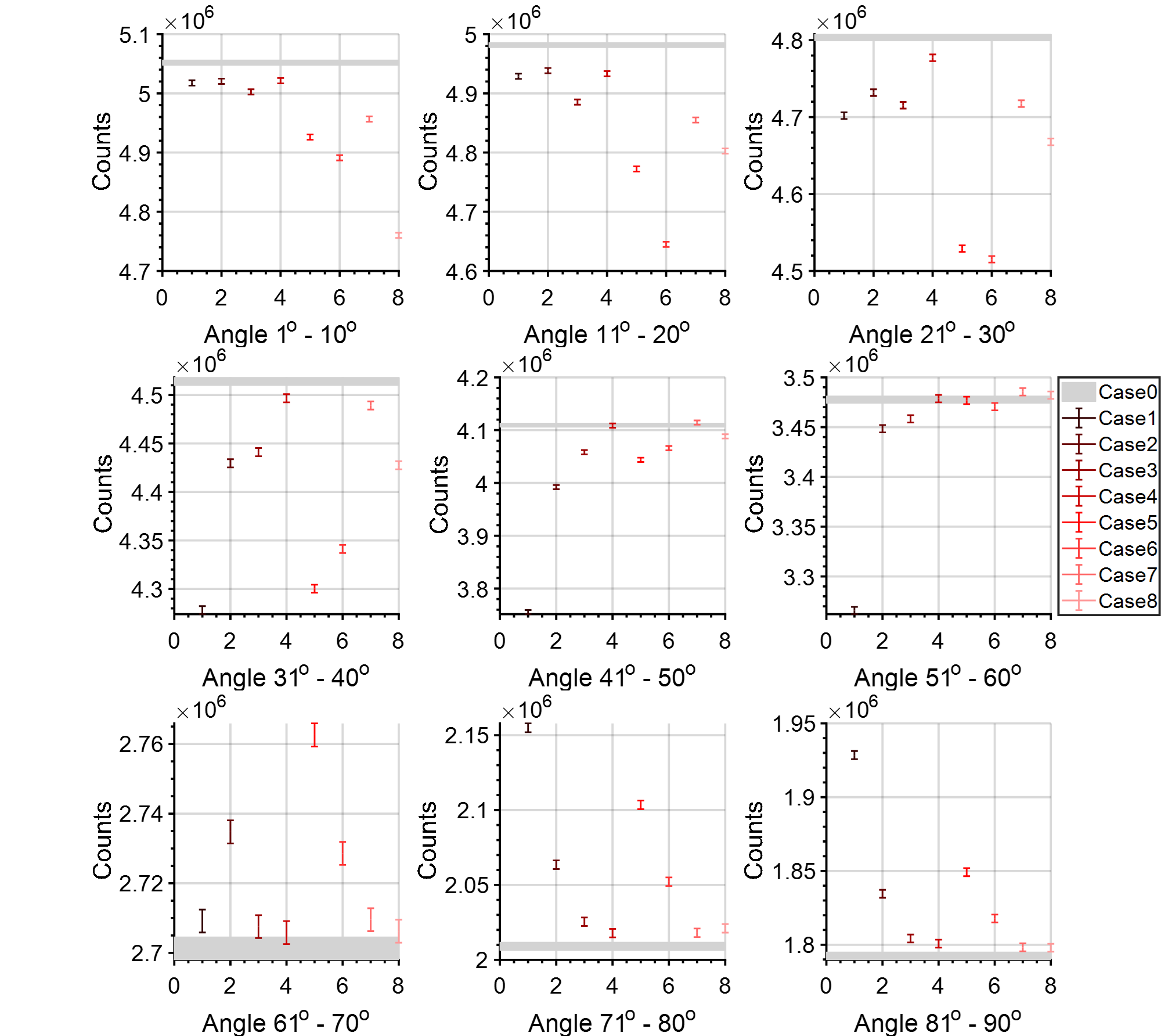}
    \caption{Angular distribution of neutron counts under $1^o$-$90^o$ for Case0-Case8 ($nps=10^8$).}
    \label{fdistribution}
\end{figure}

Fig. \ref{fdistribution} shows the variation of the neutron angular distribution with the position of the mis-loaded assembly.
The angular distribution of the neutron counts for various assembly missing cases shows that there is a statistically significant drop ($>\pm2\sigma$) of the counts compared to the fully loaded case, i.e. Case0, at the $10^o$-$30^o$ angle with respect to the neutron beam direction, for all the investigated assembly missing cases. 

\begin{figure}[h!]
    \centering
    \captionsetup{justification=centering}
    \includegraphics[scale=0.3]{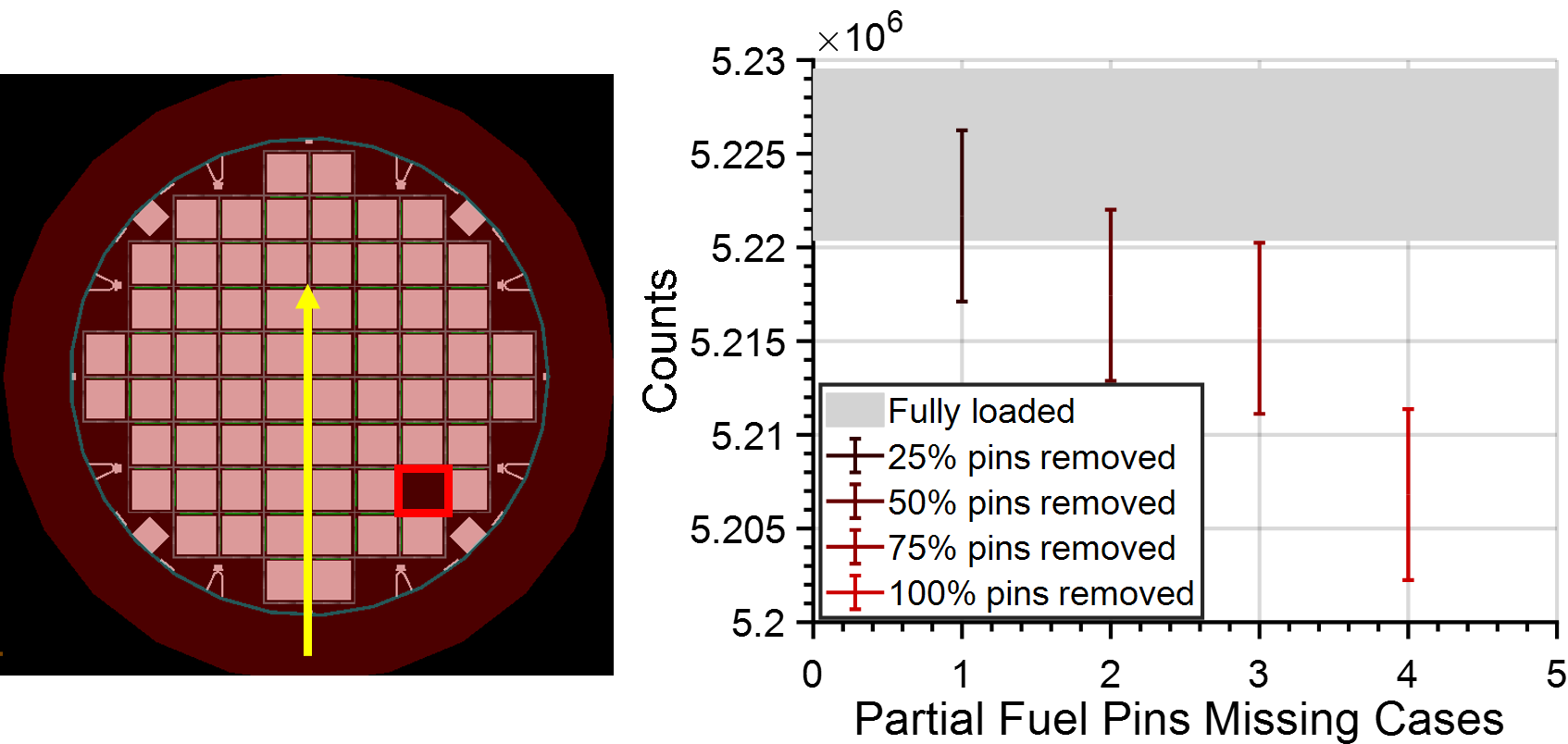}
    \caption{Neutron counts detected in the $21^o$-$30^o$ interval for the assembly on the bottom-right corner loaded with 25\%, 50\%, 75\% and 100\% of the fuel pins (the arrow in the left figure indicates the beam direction, $nps=10^8$).}
    \label{fPartialcase}
\end{figure}

We also performed the simulation with 25\%, 50\%, 75\%, and 100\% (i.e., one whole fuel assembly removed from the canister) fuel rods removed to study the sensitivity of our method to the scenarios where some of the fuel rods are removed within one fuel assembly. Fig. \ref{fPartialcase} shows the comparison of neutron counts with different partially loaded configurations for Case2. \replaced{}{The location of the partially loaded assembly is the same as Case2, as shown in Fig. \ref{fPartialcase}.} The grey bar represents the neutron counts of the fully-loaded case with an uncertainty of $\pm2\sigma$. From the results, we can identify with a confidence of $\pm2\sigma$ a partially loaded fuel assembly with a fraction of at least 75\% of the rods being removed. 

\begin{figure}[!h]

    \begin{subfigure}[c]{0.3\textwidth}
    \centering
    \includegraphics[scale=0.28]{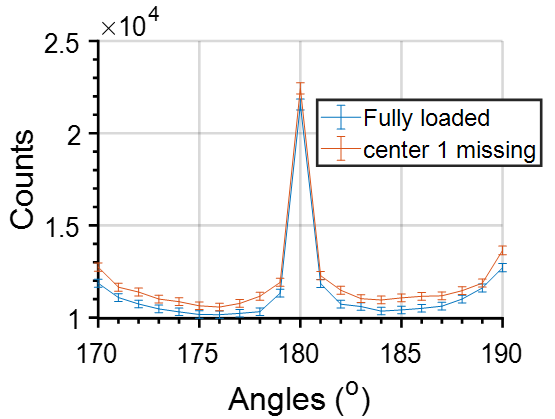}
    \caption{Incident angle: $0^o$.}
    \label{fTransmitted_na}
    \end{subfigure}
    \quad
    \begin{subfigure}[c]{0.3\textwidth}
    \centering
    \includegraphics[scale=0.28]{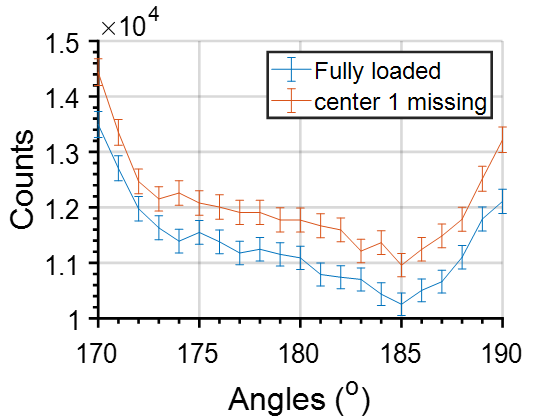}
    \caption{Incident angle: $30^o$.}
    \label{fTransmitted_nb}
    \end{subfigure} 
    \quad
    \begin{subfigure}[c]{0.3\textwidth}
    \centering
    \includegraphics[scale=0.28]{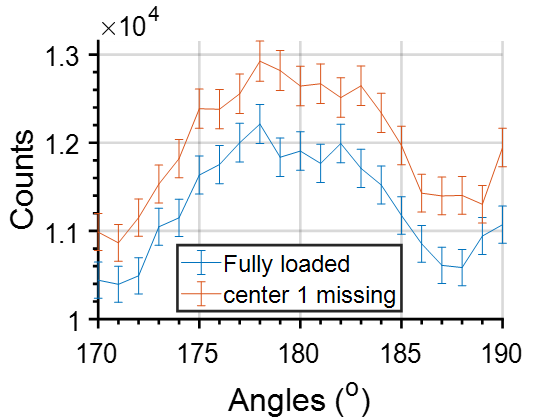}
    \caption{Incident angle: $45^o$.}
    \label{fTransmitted_nc}
    \end{subfigure}
    
\caption{Distribution of the transmitted neutrons for different incident angles (fan beam, $nps=10^8$).}
\label{fTransmitted_n}
\end{figure}

The transmitted neutron signature is shown in Fig. \ref{fTransmitted_n}. One can observe a significant difference in the neutron distribution between the single fuel assembly missing at the center region and the fully loaded cases, with an uncertainty of $\pm2\sigma$. The transmitted neutron signature can thus be used to inspect innermost bundles. However, an interrogating neutron fluence at least one order of magnitude higher (\replaced{for example, $nps=10^{10}$}{$nps=10^9$}) is needed to obtain statistically significant results, compared to the analysis based on the back-scattered neutron signature. \replaced{}{Therefore, our image reconstruction analysis, which requires the simulation of the full response matrix, was limited to the back-scattered signature.}

\subsection{Image reconstruction}
\subsubsection{Results from FISTA}
\begin{figure}[!h]
\begin{tabular}[c]{c c}
    \begin{subfigure}[c]{0.45\textwidth}
    \centering
    \includegraphics[scale=0.3]{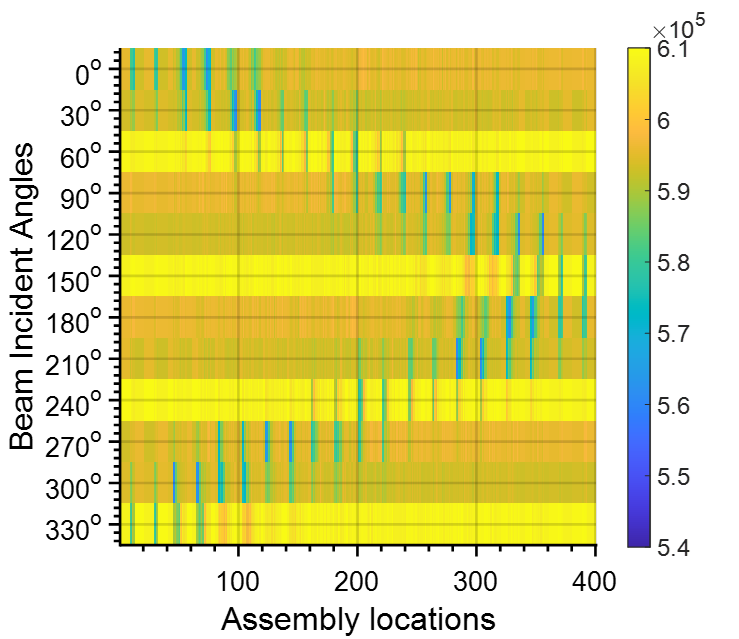}
    \caption{System response within detection angle $10^o$-$20^o$.}
    \label{f:ch4:fine_rm:a}
    \end{subfigure}
    
    \begin{subfigure}[c]{0.45\textwidth}
    \centering
    \includegraphics[scale=0.3]{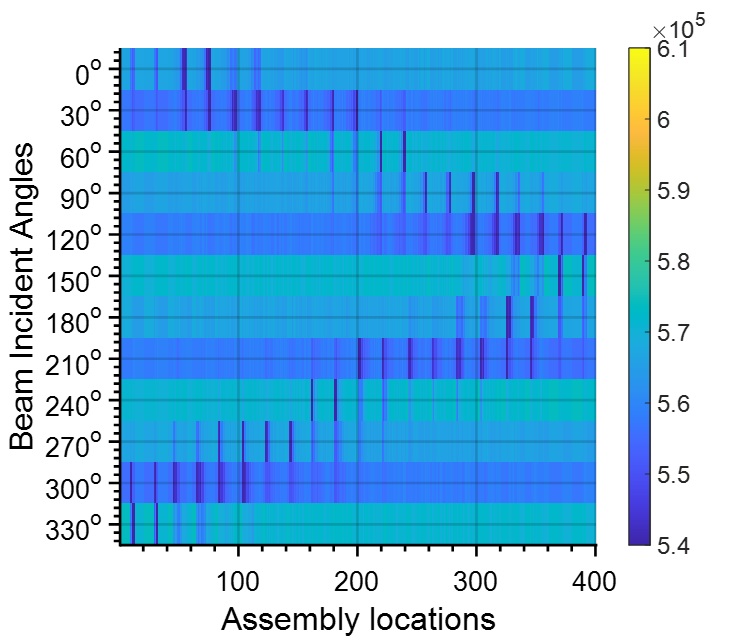}
    \caption{System response within detection angle $20^o$-$30^o$.}
    \label{f:ch4:fine_rm:b}
    \end{subfigure} 
\end{tabular}
\caption{The finer response matrix in two detection regions ($nps=10^7$\added{, the colorbar represents the response of the system to fuel rods missing in each pixel within the corresponding detection angle range}).}
\label{f:ch4:fine_rm}
\end{figure}

We used the \replaced{Geant4}{GEANT4} model to generate the response matrix, hence accurately capturing the neutron transport physics in the canister. Simulated and noise-corrupted casks with various configurations of missing fuel bundles were used to test the reconstruction algorithm.
We defined a \replaced{finer}{fine} response matrix with $N=20$ to improve the sensitivity of the reconstruction algorithm. The dimension of the finer response matrix is $M\times 20^2$, with each pixel representing the system response to one-fourth fuel assembly removed from the canister. 
Fig. \ref{f:ch4:fine_rm}a and b show the simulated fine response matrix for detection angles in the $10^o$-$20^o$ and $20^o$-$30^o$ intervals, respectively, where the interrogating beam rotated around the cask in $30^o$ intervals (y axis). 
The color bar shows the intensity of each pixel in a linear scale. From the response matrix, the intensity of the corresponding pixel drops as expected when a fuel assembly is removed. 

\begin{figure}[!h]
\begin{tabular}[c]{ccc}
    \begin{subfigure}[c]{0.3\textwidth}
    \centering
    \includegraphics[scale=0.2]{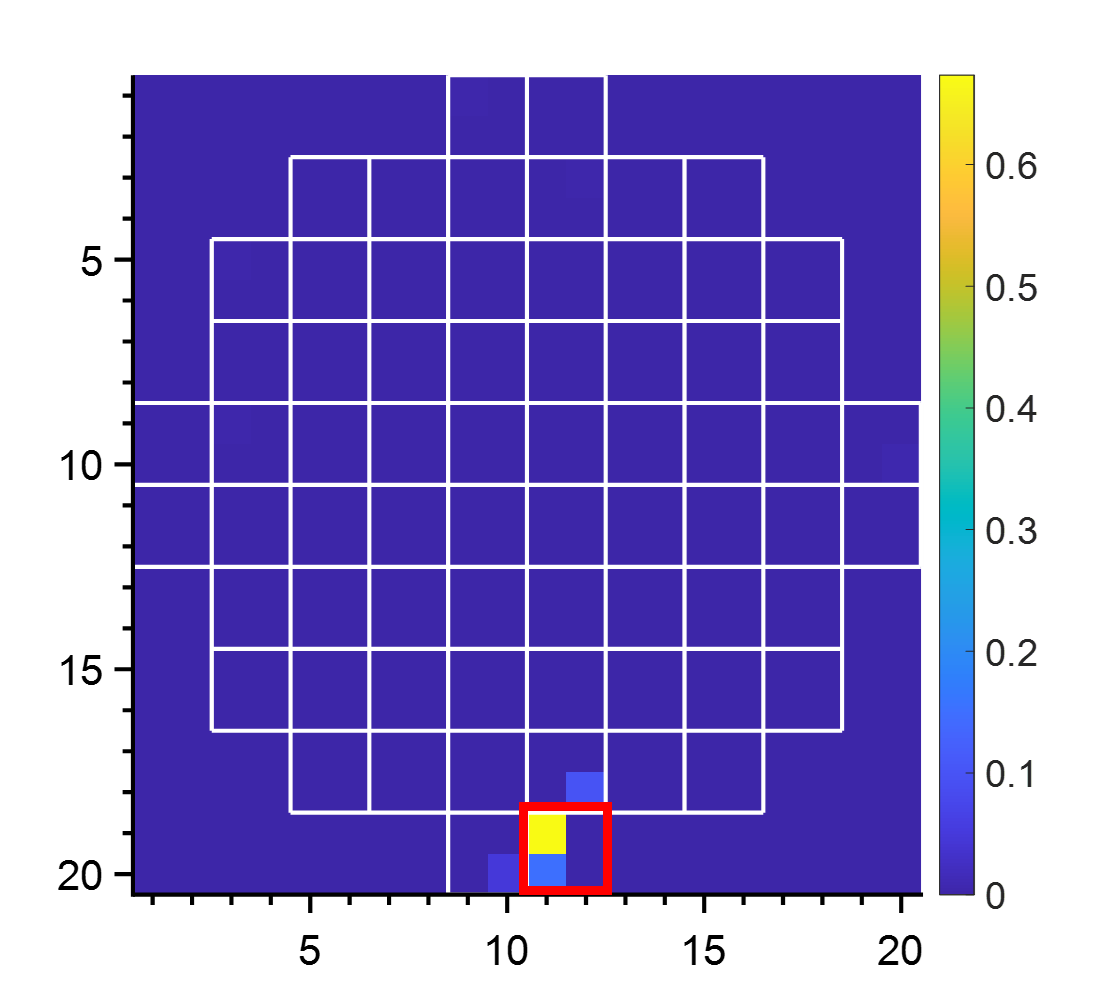}
    \caption{Case-A1.}
    \label{f:ch4:recon_images_no_noise:a}
    \end{subfigure}
    
    \begin{subfigure}[c]{0.3\textwidth}
    \centering
    \includegraphics[scale=0.2]{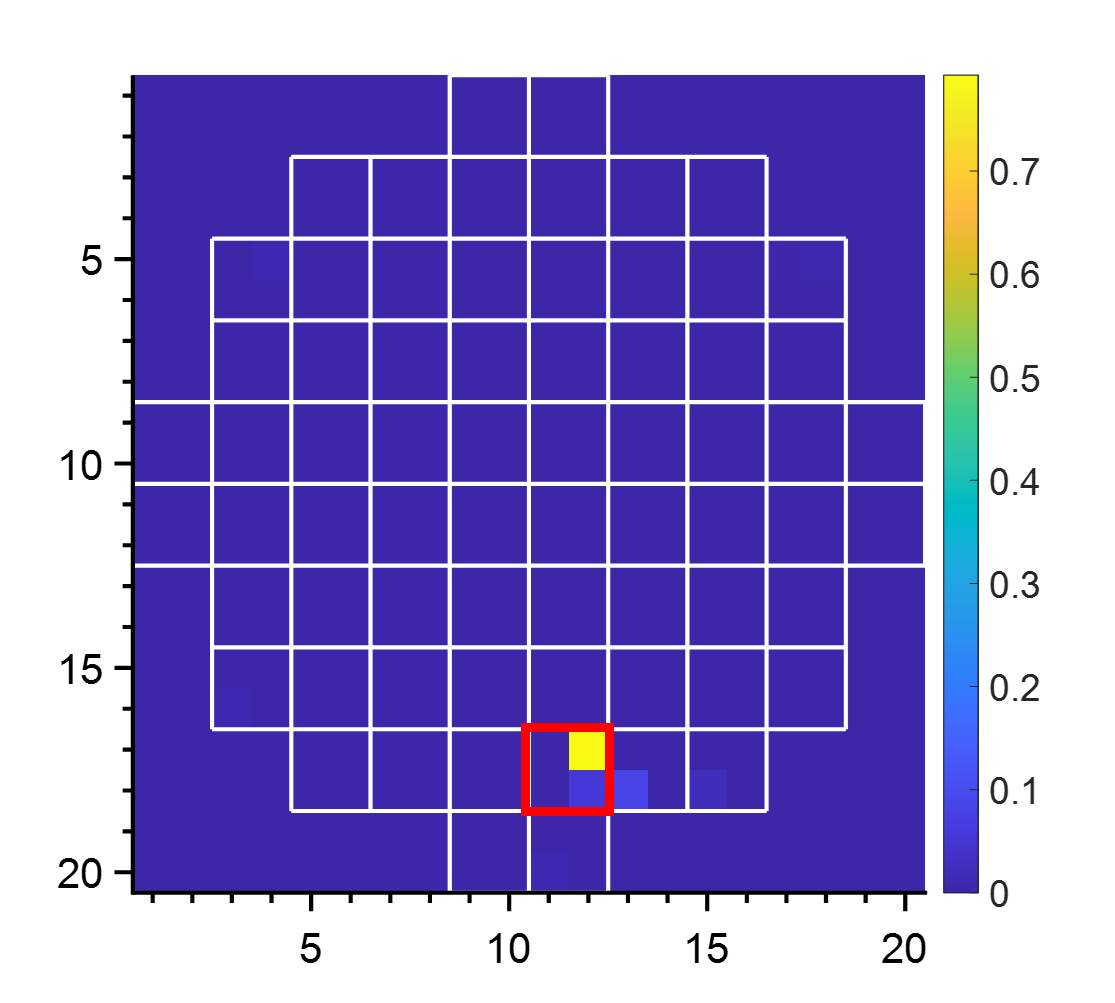}
    \caption{Case-A2.}
    \label{f:ch4:recon_images_no_noise:b}
    \end{subfigure} 
    
     \begin{subfigure}[c]{0.3\textwidth}
    \centering
    \includegraphics[scale=0.2]{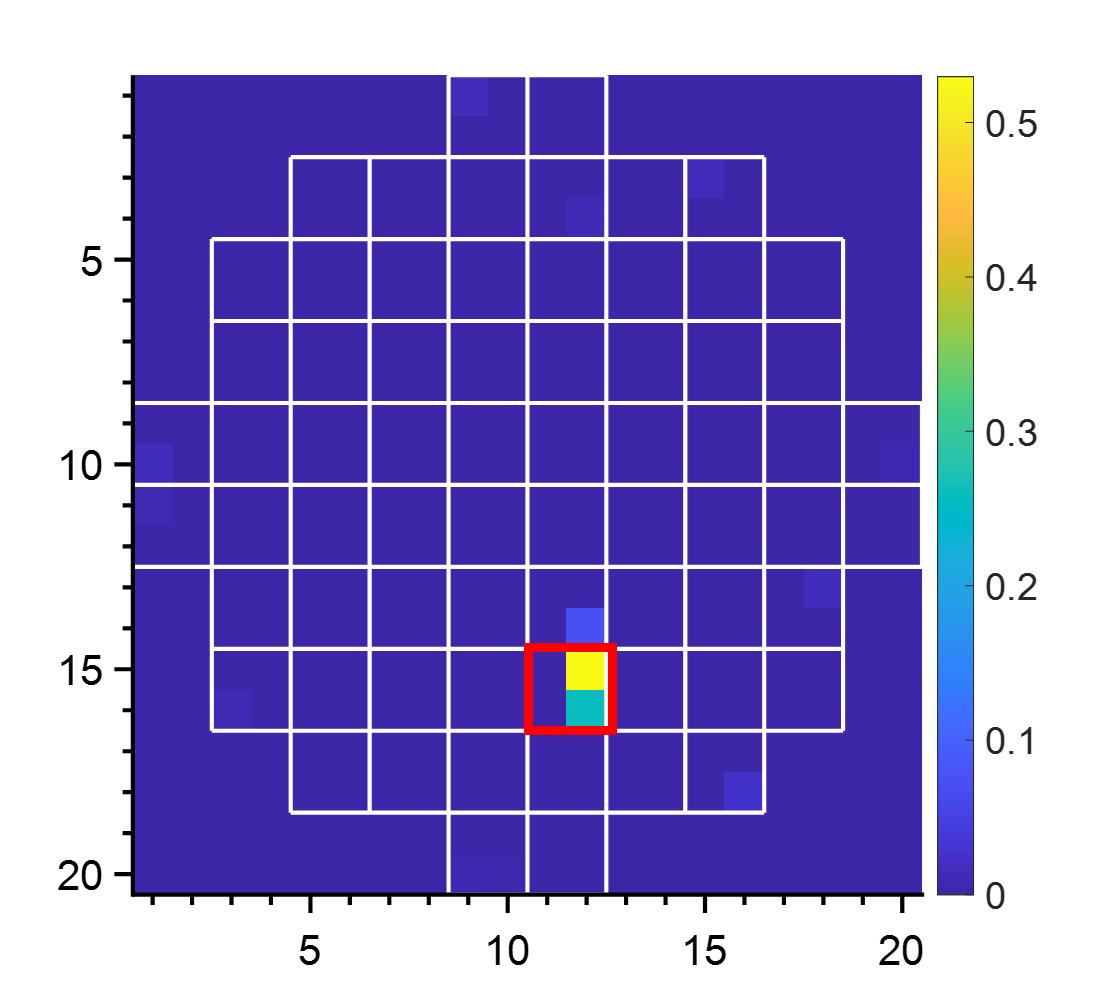}
    \caption{Case-A3.}
    \label{f:ch4:recon_images_no_noise:c}
    \end{subfigure}  \\
    
    \begin{subfigure}[c]{0.3\textwidth}
    \centering
    \includegraphics[scale=0.2]{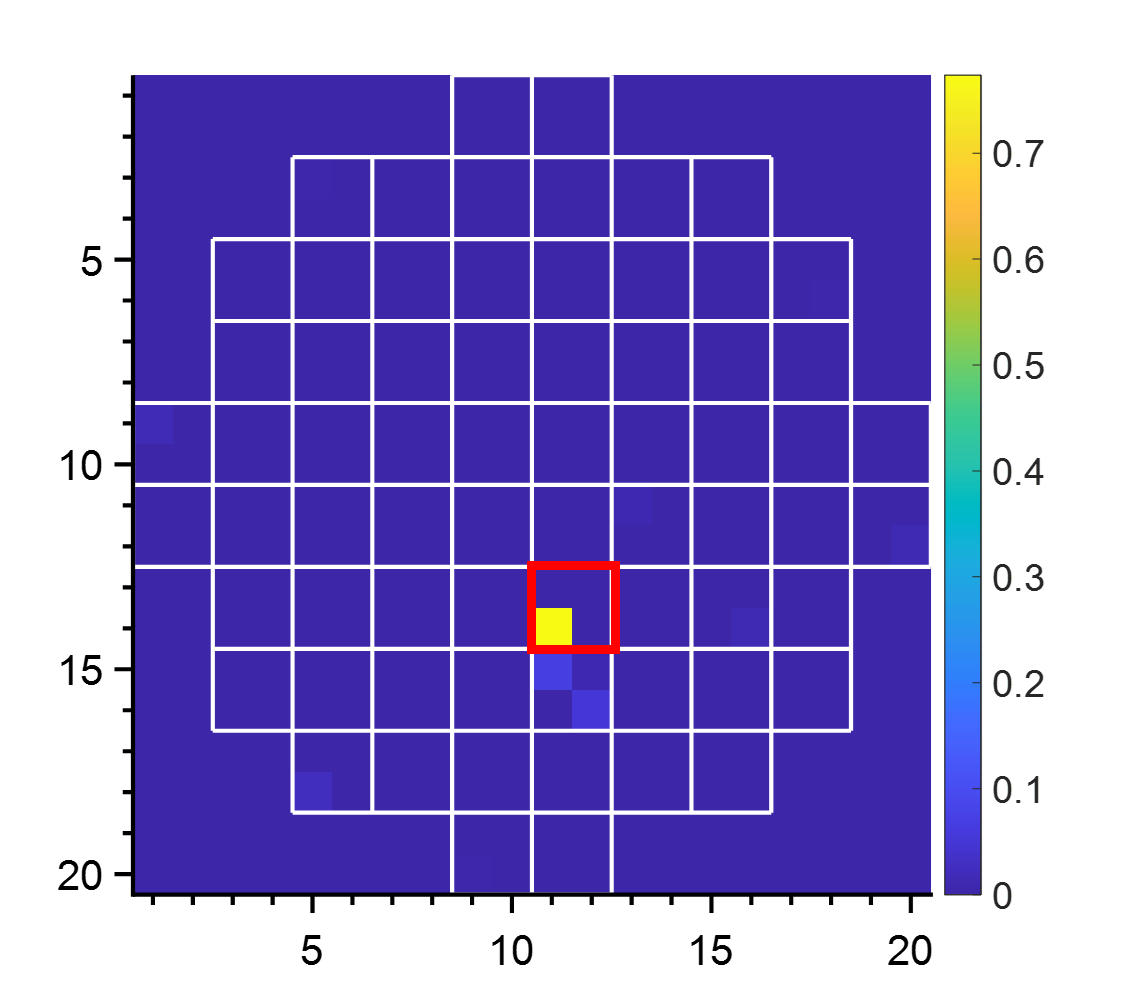}
    \caption{Case-A4.}
    \label{f:ch4:recon_images_no_noise:d}
    \end{subfigure}
    
    \begin{subfigure}[c]{0.3\textwidth}
    \centering
    \includegraphics[scale=0.2]{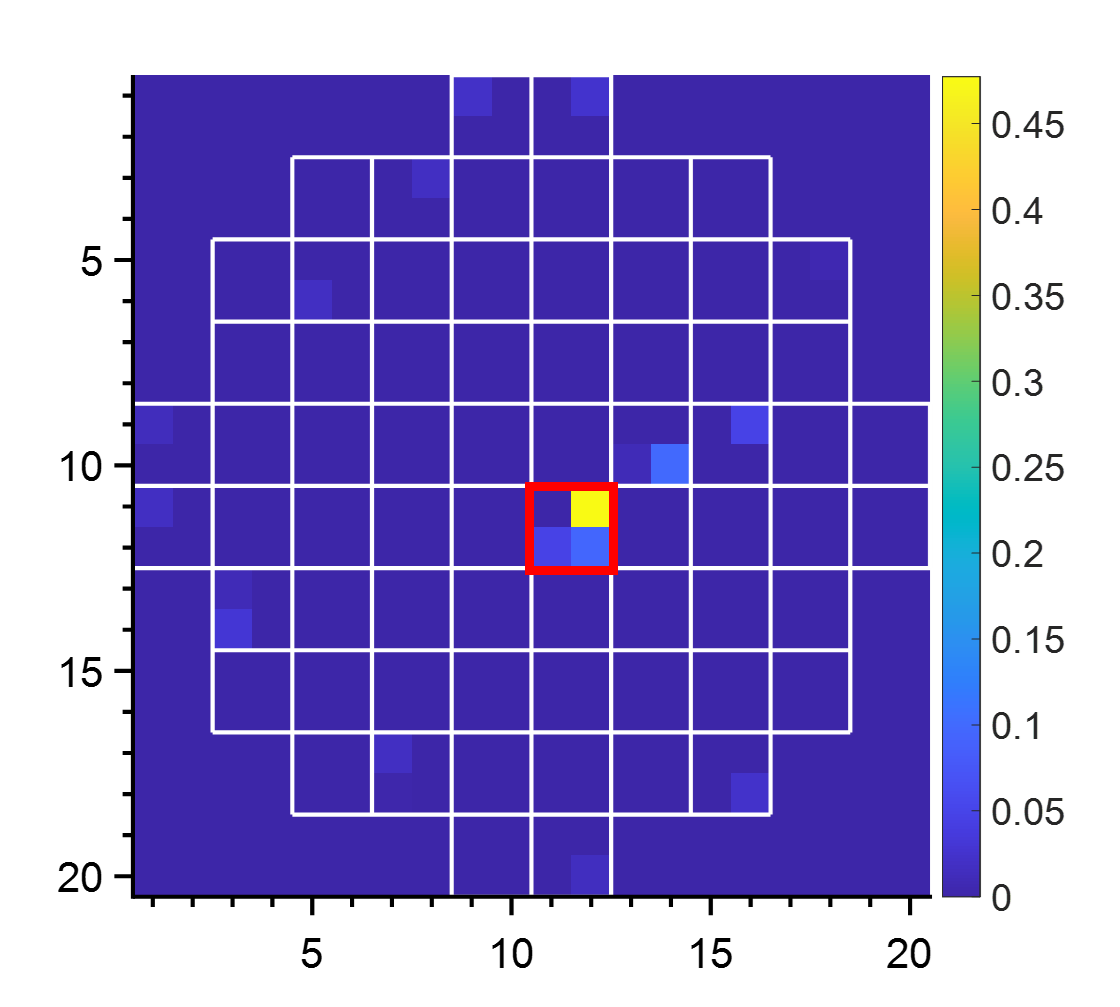}
    \caption{Case-A5.}
    \label{f:ch4:recon_images_no_noise:e}
    \end{subfigure}  \\
\end{tabular}
\caption{The reconstructed images of missing fuel bundles with changing location from the peripheral region to the center of the canister ($nps\geq10^9$added{, the colorbars represents the intensity of each pixel in the reconstructed images and an increased intensity means a possible missing fuel assembly location}). In each reconstructed image, the true location of the missing fuel assembly is highlighted by the red box.}
\label{f:ch4:recon_images_no_noise}
\end{figure}

\added{The noise associated with the measured data is due to statistical and systematical noise sources. While statistical noise depends mainly on the number of acquired counts, systematical noise sources may range from detector gain drift to bias in the source position. In our simulated model, we only considered the statistical random noise. Hence, the noise level decreases with the measurement time.} \replaced{}{Here we consider the detector with 100\% intrinsic efficiency in the 0.1 MeV - 14.1 MeV energy range.} 
\added{With a measurement time less than $10~min$ for one scan, we expect an average relative uncertainty associated with the counts at the detectors less than $\pm 0.6\%$ (1 SD). }
\replaced{}{The above estimation needs to be modified considering the practically employed detector efficiency.}
When the noise level is low (measurement time \replaced{$>10~min$}{$> 5 min$} per scanning angle), we can obtain well-reconstructed images that allow to accurately identify the location of the missing fuel assemblies, as shown in Fig. \ref{f:ch4:recon_images_no_noise}. We reconstructed several cases with one missing fuel assembly, at different locations. 
We simulated five cases named as Case-A1 to Case-A5, as shown in Fig. \ref{f:ch4:recon_images_no_noise}. \replaced{}{The readout used in the reconstruction algorithm for the five cases contains low-level of noise.} The pixel with the highest intensity corresponds to the identified assembly location by the algorithm. In these five cases, the missing fuel assembly is always correctly located, despite of its location.

\begin{figure}[!h]
    \centering
    \begin{subfigure}[c]{0.43\textwidth}
    \centering
    \includegraphics[scale=0.25]{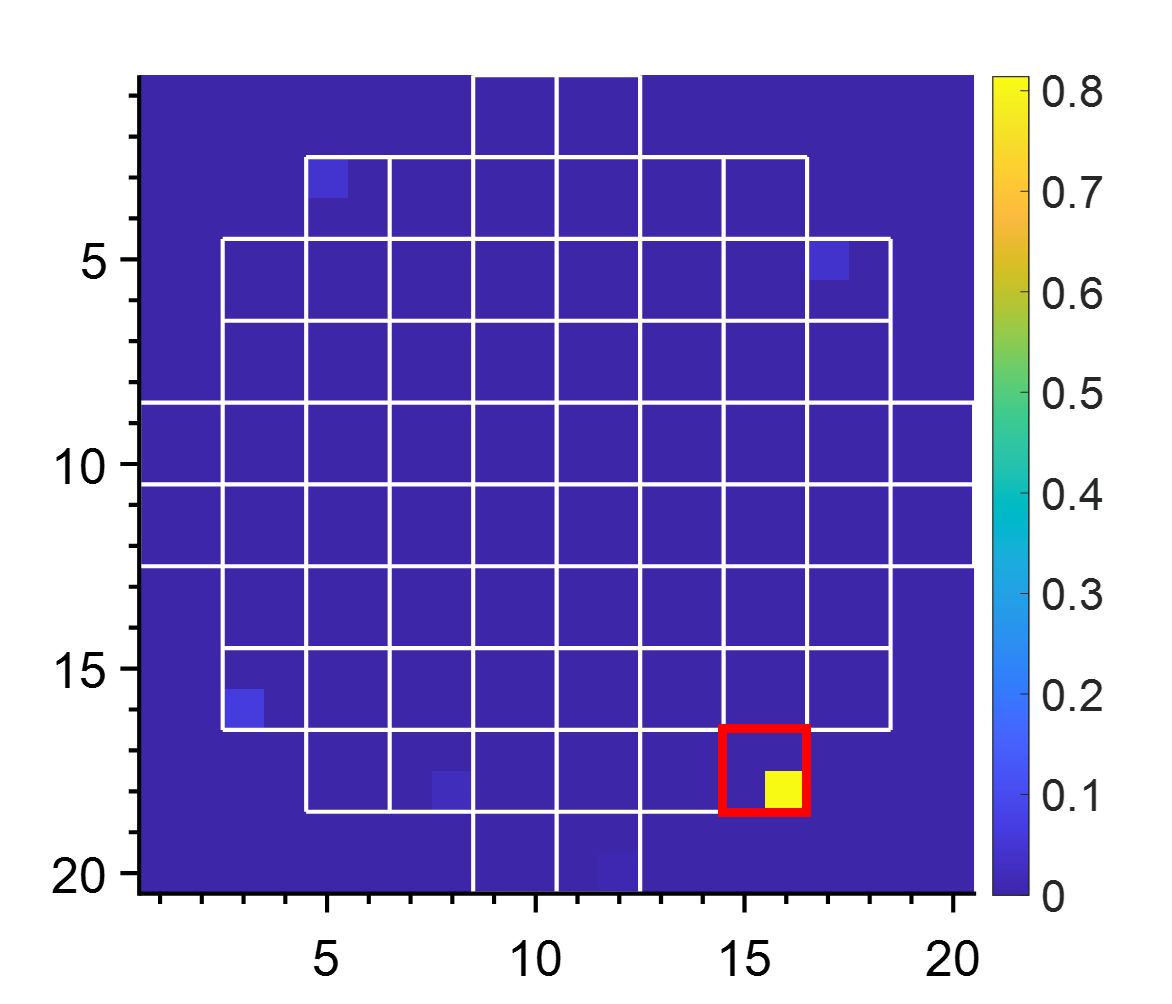}
    \caption{The reconstructed image of Case-D1.}
    \label{f:ch4:recon_img_fine_rm_diagnal:a}
    \end{subfigure}
\quad    
    \begin{subfigure}[c]{0.43\textwidth}
    \centering
    \includegraphics[scale=0.25]{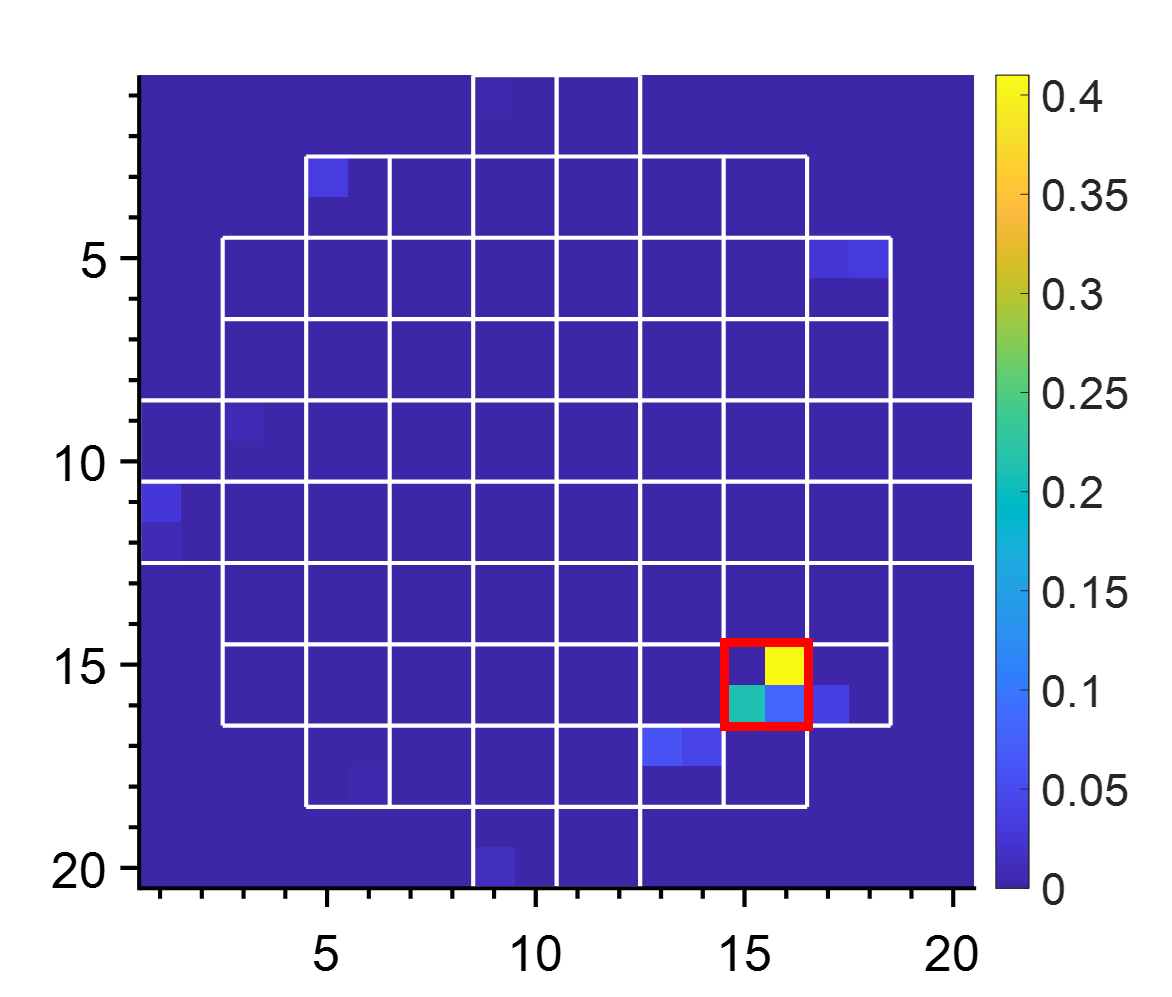}
    \caption{The reconstructed image of Case-D2.}
    \label{f:ch4:recon_img_fine_rm_diagnal:b}
    \end{subfigure} \\
    
    \begin{subfigure}[c]{0.43\textwidth}
    \centering
    \includegraphics[scale=0.25]{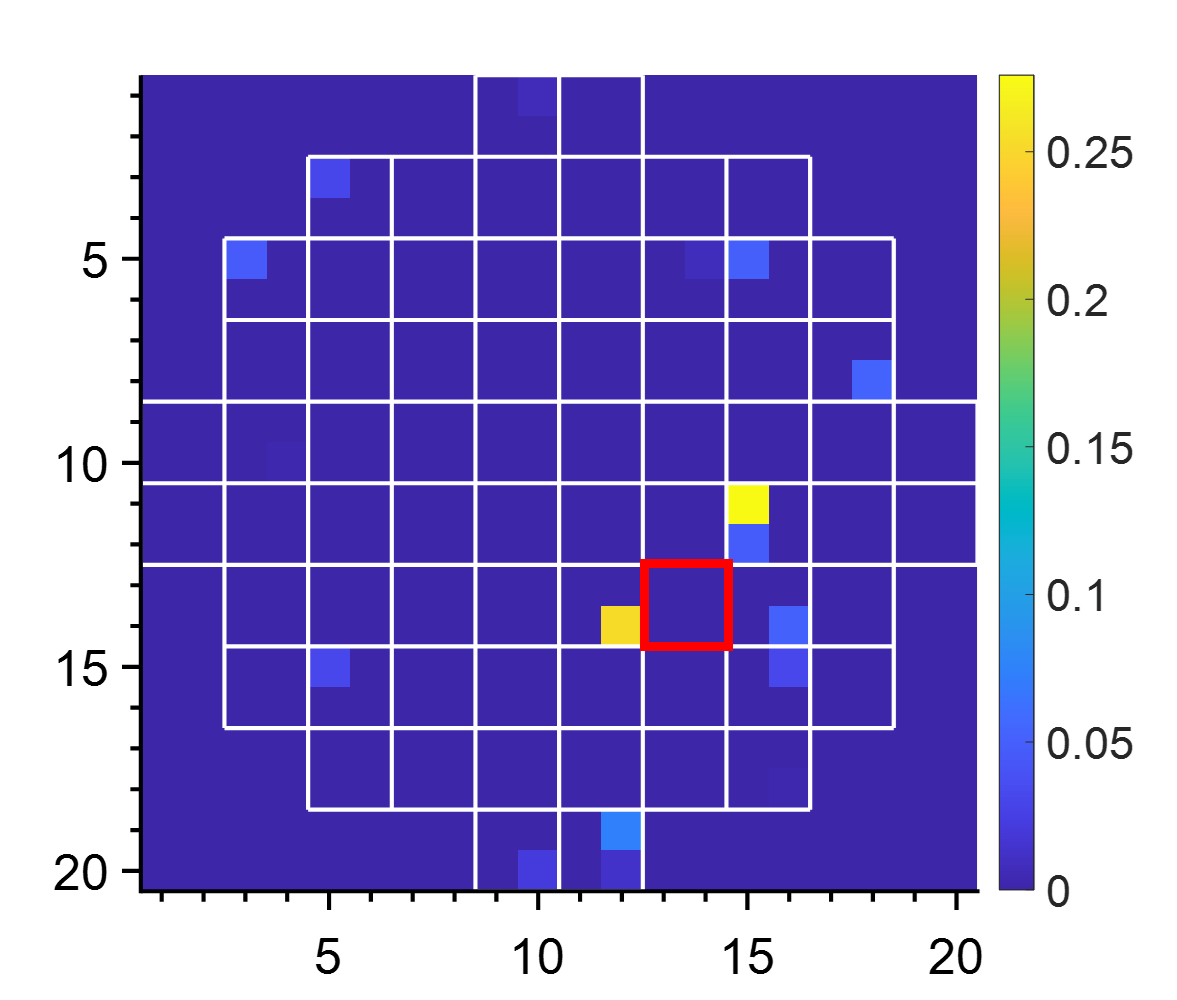}
    \caption{The reconstructed image of Case-D3.}
    \label{f:ch4:recon_img_fine_rm_diagnal:c}
    \end{subfigure}
\quad    
    \begin{subfigure}[c]{0.43\textwidth}
    \centering
    \includegraphics[scale=0.25]{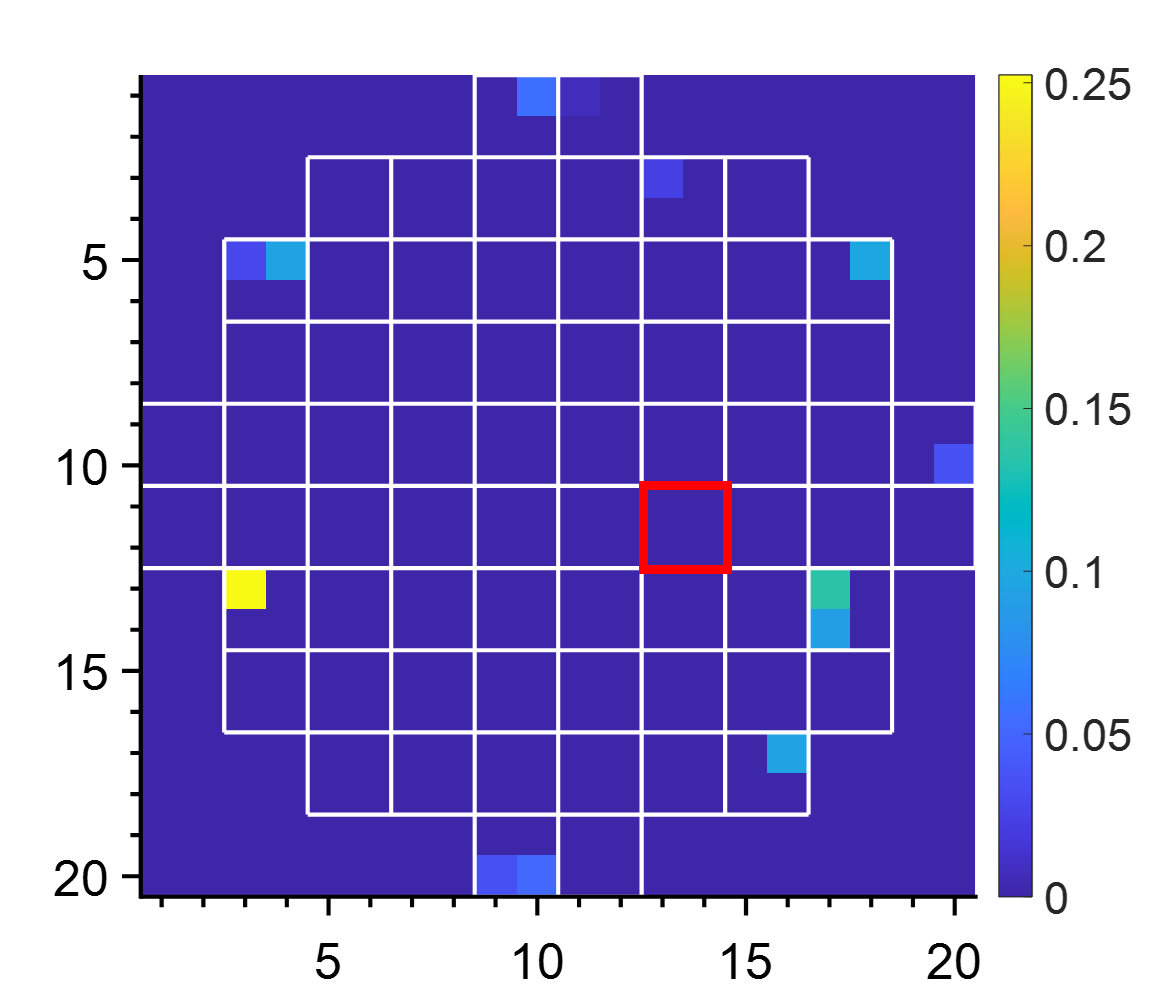}
    \caption{The reconstructed image of Case-D4.}
    \label{f:ch4:recon_img_fine_rm_diagnal:d}
    \end{subfigure}

\caption{The reconstructed images for fuel assembly with $75\%$ of pins removed along the diagonal of the canister ($nps=10^7$\added{, the colorbars represents the intensity of each pixel in the reconstructed images and an increased intensity means possible missing fuel assembly locations}).}
\label{f:ch4:recon_img_fine_rm_diagnal}
\end{figure}

We then reconstructed four cases with $75\%$ of pins removed in one fuel assembly, located along the cask diagonal to investigate the penetrability of the NDA method. 
\replaced{}{The location of the fuel assembly changes along the diagonal of the canister from the outermost layer towards the center, to investigate the sensitivity of this image reconstruction method to cask configurations with one 75\% removed assembly at various locations, i.e., the penetrability of the non-destructive assay. We named the four simulated cases as Case-D1 to Case-D4,} The reconstructed images are shown in Fig. \ref{f:ch4:recon_img_fine_rm_diagnal}.

In Fig. \ref{f:ch4:recon_img_fine_rm_diagnal}, we can successfully identify the missing fuel assemblies that are close to the sidewall of the canister. As the location of the missing fuel assembly moves from the peripheral to the center of the canister, the sensitivity of the method drops.

\begin{figure}[h!]
    \centering
    \captionsetup{justification=centering}
    \includegraphics[scale=0.3]{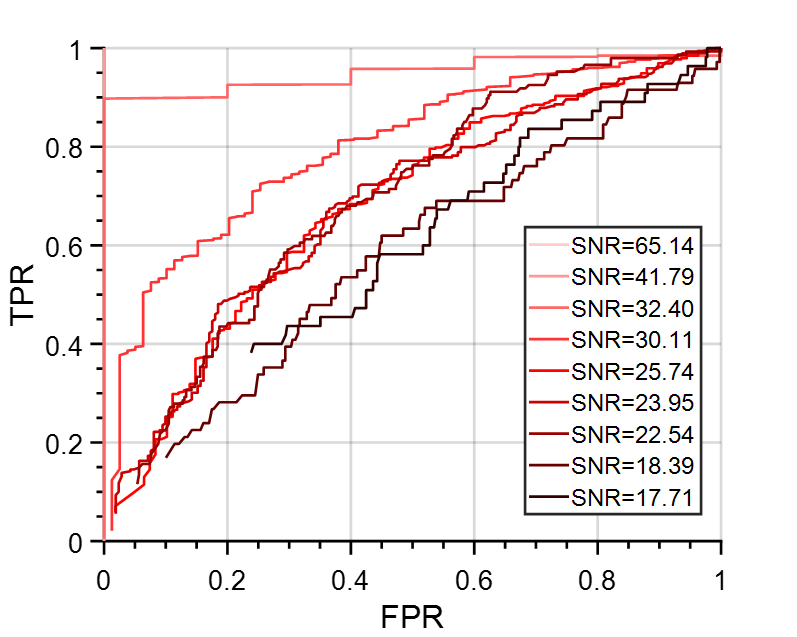}
    \caption{The ROC curves for different noise levels.}
    \label{f:ch4:ROC}
\end{figure}

We added Gaussian noise to the readout data $\mathbf{y}$ to generate noise-corrupted input data and evaluate the effect of noise on the reconstructed images. The noisy input data have an SNR ranging from 18 to 65\replaced{}{ when compared to the simulated readout data $y$}.
\replaced{}{: $SNR=65.14$, $SNR=41.79$, $SNR=32.40$, $SNR=30.11$, $SNR=25.74$, $SNR=23.95$, $SNR=22.54$, $SNR=18.39$, and $SNR=17.71$. }
We calculated the receiver operating characteristic (ROC) curve for different noise levels as shown in Fig. \ref{f:ch4:ROC}, for the case study of a missing fuel assembly at the bottom-right corner (i.e., Case7). \added{In the ROC, a fuel-missing case corresponds to a pixel intensity of the reconstructed image higher than the threshold. Conversely, a full-assembly case corresponds to a pixel intensity lower than the threshold. 
True positive (TP) cases refer to correctly identified fuel-missing cases. False positive (FP) cases refer to cases that are identified as fuel-missing whereas in reality the assemblies are full. The true positive rate (TPR) is the ratio between TP cases and all cases that are identified as fuel-missing ones. The false positive rate (FPR) is the ratio between FP cases over all cases that are identified as full-assembly ones.}
The ROC curves show the relationship of noise intensity with the accuracy of the reconstruction, that is, the higher the noise level (i.e., the lower the SNR), the worse the ability to identify the correct location of the missing fuel assembly. Therefore, to ensure the accuracy of the image reconstruction, we need sufficiently high neutron fluence (which should ensure the canister being actually irradiated by at least a total of $10^9$ source neutrons), especially for the mis-loaded scenarios with fuel assemblies located in the inner region of the canister.

\subsubsection{Results from CNN}

\begin{figure}[!h]
    \centering
    \begin{subfigure}[c]{0.43\textwidth}
    \centering
    \includegraphics[scale=0.45]{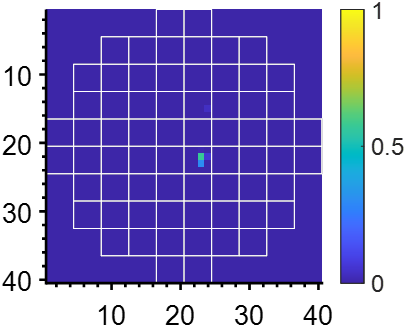}
    \caption{Input A (SNR: 3.863).}
    \label{CNN input:a}
    \end{subfigure}
\quad    
    \begin{subfigure}[c]{0.43\textwidth}
    \centering
    \includegraphics[scale=0.45]{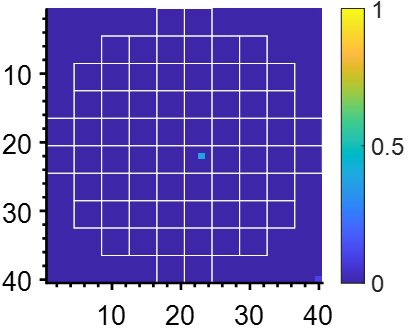}
    \caption{Output A (SNR: 9.948).}
    \label{CNN output:a}
    \end{subfigure} \\
    
    \begin{subfigure}[c]{0.43\textwidth}
    \centering
    \includegraphics[scale=0.45]{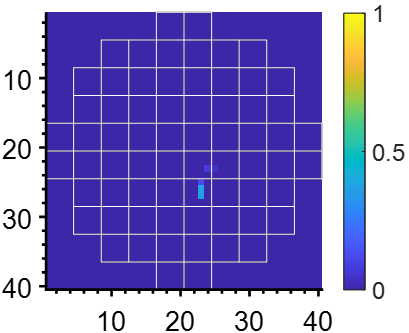}
    \caption{Input B (SNR: 1.666).}
    \label{CNN input:b}
    \end{subfigure}
\quad    
    \begin{subfigure}[c]{0.43\textwidth}
    \centering
    \includegraphics[scale=0.45]{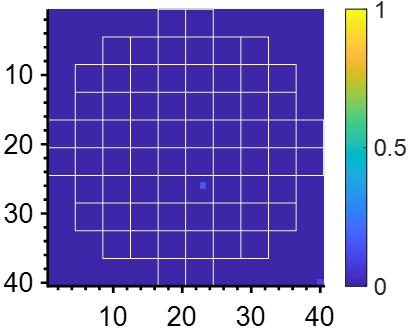}
    \caption{Output B (SNR: 2.693).}
    \label{CNN output:b}
    \end{subfigure}

\caption{The CNN output for inner fuel assembly missing cases using the noisy reconstructed images from FISTA as input. \replaced{The colorbars are normalized to 1 based on the corresponding ground truth where the intensity is set as 1 for the missing fuel assembly pixel}{The color bars are normalized to 1}.}
\label{CNN_outputs}
\end{figure}

\added{
The images reconstructed with FISTA tend to be noisy when the missing assembly is located in the center part of the cask. Hence, we applied the CNN model to improve the quality of the reconstructed images.

Fig. \ref{CNN_outputs} shows the noisy reconstructed images from FISTA as the network inputs and the corresponding CNN outputs. From the comparison between the inputs and corresponding outputs, the SNR of the reconstructed images can be improved by a factor of at least 1.5.
}


\section{Discussion and Conclusions}

We presented the results of a simulation-based study aimed at identifying suitable operating parameters for the non-destructive inspection of SNF with 14.1~MeV interrogating neutrons.
We simulated a \replaced{D-T}{DT} neutron generator with a source strength of \replaced{10$^{10}$}{10$^9$} n/s, suitably collimated to minimize off-axis irradiation that would otherwise increase the measurement background without adding useful information. Fast neutrons emitted as a result of the interactions of the primary beam with the SNF were detected and used as a signature to identify missing fuel rods. 
We found that a direct angle-differential fast neutron measurement is a viable approach to detect missing pins inside a fuel cask. \replaced{Based on the proposed method, one assembly with at least 75\% pins (69 out of 92 fuel pins) removed can be identified. The detection of a single missing pin is challenging with this method. Other time-correlated signatures could be considered in the future to improve the specificity of the NDA method to fissile and fissionable material.
For example, longer irradiation time would enable the measurement of coincidence neutrons that are a specific signature of induced fission.}{It is possible to identify a partial loading of at least $75\%$ in a $5 min$ measurement time, for an assembly located in any canister in the two outermost layers of the SNF cask.} 
\added{In general, a longer inspection time, corresponding to a neutron fluence irradiating the cask  $\geq10^9$, would yield a lower relative uncertainty of the neutron counts, which could improve the overall NDA performances.}

\replaced{}{Fast-neutron sensitive detectors, such as organic scintillators, He-4 proportional counters, fast-neutron sensitive fission chambers, or moderated thermal neutron detectors Thermal neutrons were disregarded and} Most of the fast neutrons were produced by scattering reactions while fissions accounted for less \replaced{than}{that} $0.1\%$ of all the interactions occurring within the cask. These conditions allowed us to adopt an iterative linear imaging reconstruction approach based on the accurate simulation of the system response to missing fuel located throughout the cask. We hence reconstructed the bi-dimensional cross-section of a SNF assembly by solving an inverse problem with FISTA. When solving the inverse problem, the location of the missing fuel can be reconstructed by unfolding it from the neutron angular distribution and the knowledge of the system response to missing bundles within the entire SNF cask field-of-view. 
\replaced{}{We processed  machine-learning based methods are being implemented to combine the back-scattered neutron signature and the transmitted neutron signature to increase the image sensitivity and study its dependence on the presence of partial defects, i.e. depleted fuel pins replaced with materials such as iron or lead. }\added{Images reconstructed using FISTA tend to be noisy when the missing fuel assembly is located in inner region of the canister. We denoised the reconstructed images and improved the SNR by applying a CNN-based model. By optimizing the parameters of the CNN architecture, such as the \replaced{}{the }number of layers, number of convolutional filters, the filter size, the batch size, and the number of training epochs, we were able to identify and locate one missing fuel assembly in the center of the canister, hence improving the overall penetrability of the method, with respect to the direct measurement of the back-scattered signature.} \replaced{}{We set the number of epochs for the training process to minimize the loss function, i.e., the L2 norm between the network output and the ground truth.}

This method allowed us to accurately locate the position of a \added{full} missing bundle with an overall measurement time \replaced{less than 2 hours}{of 1 hour} and a source strength of \replaced{at least $10^{10}$}{$10^9$} n/s in the 4$\pi$ solid angle, using organic scintillation detectors.

\added{
Using organic scintillation detectors to perform the tomographic inspection, the measurement time of a single scan is approximately $8.5~min$ to achieve an average relative uncertainty associated with the counts at the detectors less than $0.6\%$. Therefore, the measurement time for an inspection including 12 scans was estimated to be 1.7 hours.}

\added{The simulated inspection time of less than two hours is much shorter when compared to other proposed methods, which may require an inspection time longer than 24 hours \cite{poulson2017cosmic}. 
The estimated inspection time can be further reduced using a stronger source and a closer source-to-canister distance. However, reducing the source-to-canister distance will require further collimation of the source that might reduce the field-of-view of the source and the overall neutron flux irradiating the assemblies. }
\replaced{}{There are some limitations about the proposed methods, for example, the proposed imaging method works well only for the peripheral bundles and the estimation of the inspection time less than 2 hours should be corrected by the actually employed detector efficiency factors.}

\added{The overall scope of this work was to study the feasibility of neutron imaging of spent fuel through simulations. A realistic experimental verification of the neutron tomographic imaging of the canister was not conducted on SNF casks. 
We are currently validating our simulated model using a small-scale experiment based on a P385 neutron generator. 
}

\section*{Acknowledgements}
This work is funded in part by the Department of Energy (DOE) under contract number 000128931. The Argonne Leadership Computing Facility (ALCF) provided the access to super-computing resources.

\newpage
\section*{References}
\typeout{}
\bibliography{3references}

\end{document}